\documentclass[12pt,preprint]{aastex}
\usepackage{psfig}
\def\la{\mathrel{\mathpalette\fun <}}

\def\fun#1#2{\lower3.6pt\vbox{\baselineskip0pt\lineskip.9pt
        \ialign{$\mathsurround=0pt#1\hfill##\hfil$\crcr#2\crcr\sim\crcr}}}
\def\ec{{\cal E}}
\def\kpc{\,{\rm kpc}}
\def\mld{\left (M/L_{\rm R}\right )_{\rm d}}
\def\mlb{\left (M/L_{\rm R}\right )_{\rm b}}

\begin{document}

\title{Equilibrium Disk-Bulge-Halo Models for the Milky
Way and Andromeda Galaxies}

\author{Lawrence M. Widrow\altaffilmark{1}}
\affil{Department of Physics, Queen's University, 
Kingston, ON, K7L 3N6, Canada}
\altaffiltext{1}{widrow@astro.queensu.ca}

\author{John Dubinski\altaffilmark{2}}
\affil{Department of Astronomy and Astrophysics, University of 
Toronto, 60 St. George Street, Toronto, ON, M5S 3H8, Canada}
\altaffiltext{2}{dubinski@astro.utoronto.ca}

\begin{abstract}

We describe a new set of self-consistent, equilibrium disk galaxy
models that incorporate an exponential disk, a Hernquist model bulge,
an NFW halo and a central supermassive black hole.  The models are
derived from explicit distribution functions for each component and
the large number of parameters permit detailed modeling of actual
galaxies.  We present techniques that use structural and kinematic
data such as radial surface brightness profiles, rotation curves and
bulge velocity dispersion profiles to find the best-fit models 
for the Milky Way and M31.  Through N-body realizations of
these models we explore their stability against the formation of bars.
The models permit the study of a wide range of dynamical phenomenon
with a high degree of realism.

\end{abstract}



\section{Introduction}

The modeling of spiral galaxies to match photometric and dynamical
measurements is a time-honoured endeavor.  A complete and accurate
characterization of the structural and kinematical properties of
galaxies is necessary to understand the diversity of galactic dynamical
behaviour
and the origin of galaxies in the current cosmological paradigm.

This paper presents a new set of models for the phase-space
distribution functions (DFs) of axisymmetric disk galaxies.  The
models consist of an exponential disk, a Hernquist model bulge
\citep{her90}, an NFW halo \citep{nfw96}, and a central, supermassive
black hole.  They are defined by a large number of parameters which
permit detailed modeling of real galaxies by fitting to observational
constraints.  Though the models represent self-consistent equilibrium
solutions to the coupled Poisson and collisionless Boltzmann (CB)
equations, they are subject to both local and global nonaxisymmetric
instabilities and are therefore suitable as initial conditions for
numerical studies of galactic dynamics.

The standard practice in modeling disk galaxies has been to assume
simple functional forms for the space density and gravitational
potential of the disk, bulge, and halo and then fit to a wide range of
observational data.  For the Milky Way galaxy, optimal structural
parameters for these ``mass models'' are found primarily from surface
brightness photometry, local stellar kinematics, the circular rotation
curve, and observations of dynamical tracer populations such as
globular clusters, the Magellanic clouds and the dwarf
satellite galaxies.  These observational constraints have been used to
determine the best-fit model parameters based on $\chi^2$-minimization
techniques.  (See, for example, \citet{inn73}, \citet{clu77},
\citet{bah80}, \citet{cal81}, \citet{kui91}, \citet{roh88},
\citet{mal95}; \citet{koc96}, \citet{deh98a}; \citet{eva99},
\citet{kly02}.)

For external galaxies, the problem is somewhat simpler because of the
advantage of an outside view and the smaller number of observational
constraints.  Optical and infrared de-projected surface brightness
profiles of the stellar distribution are combined with rotation curves
measured from HI gas kinematics to build mass models.  The standard
method is to perform a bulge-disk decomposition of the surface
brightness profile usually into an $R^{1/4}$-law surface density
profile for the bulge and a radial exponential profile for the disk.
(See, for example, \citet{ken85}, \citet{sim86}, and \citet{bin98};
see also \citet{cou96} and references therein for a discussion of
bulge-disk decomposition with the more general S\'{e}rsic law for the
bulge.)  The halo model parameters are inferred by fitting
multi-component mass models to the rotation curve data assuming values
for the mass-to-light ratios of the disk and bulge stars.  For nearby
galaxies such as M31, kinematic data for the globular cluster,
planetary nebulae and satellite systems are used to refine the models
\citep{eva00}.

There has also been parallel work in developing realistic N-body
realizations of galaxy models for numerical experimentation in disk
stability and galaxy interactions and mergers.  To generate an N-body
realization, it is necessary to model the full DFs for the
multi-component system, in principle by solving the coupled Poisson
and CB equations.  Owing to the complexity of these equations, various
approximation schemes have been used.  Early work on disk galaxy
simulations set up initial conditions by estimating the velocity
dispersion for a stable disk model with Toomre parameter $Q>1$ and
then assuming a locally Gaussian distribution for the velocities (see,
for example, \citet{sel85}).  \citet{bar88} built 3-component models
by adiabatically growing a disk potential within a live halo model and
then adding an N-body disk with initial conditions generated as above.
\citet{her93} presented a simpler approximate presciption for
3-component disk-bulge-halo models where the bulge and halo velocity
distributions are assumed to be Maxwellian, truncated at the local
escape speed and with dispersions estimated from the Jeans equations.
These methods, while convenient, produce models that are slightly out
of equilibrium and so require some relaxation time to damp out
transients.  Moreover, the models readjust to a state different from
the one proposed (see, for example, \citet{kaz04}).

As numerical methods and resolution improve it has become necessary to
develop more sophisticated techniques to generate initial conditions.
Toward this end, \citet[][hereafter KD]{kui95} presented a set of
semi-analytic models for the DFs of disk galaxies consisting of an
exponential disk, a centrally concentrated bulge, and an extended
halo.  The DFs are constructed from integrals of motion and represent
equilibrium solutions to the Poisson and CB equations to a very good
approximation and they have been used extensively to study different
aspects of galaxy internal dynamics and interactions
(e.g. \citet{dub95}, \citet{dub98a}, \citet{dub99}, \citet{gar02},
\citet{one03}, \citet{wid03}).

Fitting models to actual galaxies requires one to ``observe'' the
model and compare with real observations.  By providing the full DFs,
the KD models enable one to add a level of realism to these
pseudo-observations that is not possible with simple mass models.  For
example, a mass model rotation curve is constructed directly from the
potential rather than from model line-of-sight velocities.  With the
KD models it is possible to construct the stellar rotation curve
directly from stellar velocities and thereby incorporate asymmetric
drift.  Gravitational microlensing experiments provide another example
where improved pseudo-observations are possible since the full DF
allows for a self-consistent calculation of the predicted event rate
distribution \citep{wid03}.

The bulge and halo of the KD models are characterized by a constant
density core.  Since their development it has become widely accepted
from numerous cosmological simulations that dark halos have cuspy
centers \citep{dub91, nfw96}.  Detailed analysis of the central
surface brightness and velocity dispersion profiles of early type
galaxies and bulges suggest that these systems are also cuspy at their
centres and furthermore contain supermassive black holes.
We are therefore motivated to develop a new set of axisymmetric disk
galaxy models with cuspy halos and bulges that also allow for the
self-consistent addition of a supermassive central object.  These
models have many purposes.  They provide the gravitational potential
and mass and velocity distributions from the sphere of influence of
the central black hole in the inner few parsecs out to the virial
radius of the model galaxy.  Circular velocity curves, rotation curves
incorporating asymmetric drift, velocity ellipsoids, line-of-sight
velocity distributions (LOSVDs), and surface density (brightness)
profiles can be generated easily.  The large parameter space permits a
wide range of models for comparison to real galaxies and the good
quality of the initial equilibria make them ideal for studying subtle
dynamical processes such as bar formation and disk warping and
heating.

The DFs are described in Section 2.  Examples that match photometric
and dynamical data for the Milky Way and M31 are presented in Section
3.  This section also presents results from numerical experiments on
the stability of these models as well as a brief discussion of an M31
model that incorportates a supermassive black hole.  We conclude in
Section 4 with a summary and a discussion of possible applications of
the models.


\section{Distribution Functions}
\label{sec:DBHModels}

The phase-space DFs for the disk, bulge, and halo of the KD models are
chosen analytic functions of the integrals of motion.  By Jeans
theorem, any such DF yields a steady-state solution of the CBE in any
potential that respects these integrals \citep{bin87}.  A self-consistent
self-gravitating model is one in which the potential and space density
also satisfy the Poisson equation.

The KD models are, by design, axisymmetric with two known integrals of
motion, the energy $E$ and the angular momentum about the symmetry
axis $J_z$.  KD choose the King model DF for the bulge which, in
isolation, yields a system that is spherically symmetric and has a
density profile characterized by a constant density core, an $r^{-2}$
fall-off at intermediate radii, and a finite ``tidal'' radius where
the density vanishes.  For the halo, KD use the DF of a lowered Evans
model which also exhibits a constant density core, power law
intermediate region, and tidal radius.  (We denote the tidal radius of
the halo, which represents the outer edge of the system as a whole, as
$r_t$.)

The configuration and velocity space distributions of the King and
lowered Evans models are modified once they are incorporated into a
multi-component model.  Recall that for a system of collisionless
particles in a static potential, any $f(E)$ (i.e., any DF that is a
function only of the energy) yields a steady-state solution of the CBE
whose velocity distribution is isotropic \citep{bin87}.  For an
isolated self-gravitating system, $\rho$ and $\Psi$ are necessarily
spherically symmetric.  (For a proof, see \citet{per96}.)  In the
presence of an aspherical external potential, an $f(E)$ will yield an
aspherical mass distribution through the dependence of $E$ on $\Psi$.
However, the velocities remain isotropic so long as $f$ does not
depend on angular momentum or some other integral of motion.  For the
multi-component models considered here and in \citet{kui95}, the disk
potential causes a slight flattening of the bulge and halo.

In this section we describe new models that have an NFW halo, a
Hernquist bulge, and a central supermassive black hole.  We begin by
presenting DFs for the bulge and halo taken in isolation and then
describe how these DFs are modified for composite models.

\subsection{The halo distribution function}

\citet{nfw96} found that the density profiles of dark matter halos in
their cosmological simulations have a ``universal'' shape (the
so-called NFW profile) of the form

\begin{equation}
\label{eq:nfw}
\rho_{\rm NFW}(r) = 
\frac{\rho_h}{\left (r/a_h\right )
	\left (1 + r/a_h\right )^2}
\end{equation}

\noindent where $a_h$ is the scale radius,
$\rho_h\equiv\sigma_h^2/4\pi a_h^2$ is a characteristic density, and
$\sigma_h$ is a characteristic velocity dispersion.  (Here and
throughout we set Newton's constant $G=1$.)  In contrast with the
profile of the Evans model halo, $\rho_{\rm NFW}$ has an inner
$r^{-1}$ cusp and an extended $r^{-3}$ outer halo.  The gravitation
potential for this profile is

\begin{equation}
\label{eq:nfwpotential}
\Phi_{\rm NFW}(r) = 
-\sigma_h^2\,\frac{\log{\left (1 + r/a_h\right )}}{r/a_h}~.
\end{equation}

Our strategy is to use a DF that, in the absence of a disk or bulge,
yields a spherically symmetric NFW halo.  We assume that the velocity
dispersion is isotropic so that the DF depends only on the energy $E$.
(Formally, $E$ is the energy per unit mass.)
For a given density profile, the DF can then be calculated through an
Abel transform \citep{bin87} a procedure that has been carried out for
the NFW profile by \citet{zha97}, \citet{wid00}, and \citet{lok00}.
We write the DF as a function of the relative energy $\ec\equiv -E$

\begin{equation}
\label{eq:nfw_df1}
f_{\rm NFW}\left (\ec\right ) =
	\sigma_h^{-1} a^{-2}{\cal F}_{\rm NFW}
	\left (\ec/\sigma_h^2\right )
\end{equation}

\noindent 
For the dimensionless DF $\cal F$ we use the analytic fitting formula

\begin{equation}
\label{eq:dimensionlessDF}
{\cal F}_{\rm NFW}\left (\epsilon\right )=\left\{\begin{array}{ll}
\frac{3}{2^{9/2}\pi^2}\epsilon^{3/2}\left (1-\epsilon\right )^{-5/2}
\left (\frac{-\log{\epsilon}}{1-\epsilon}\right )^\alpha e^P
		& \mbox{for $0<\epsilon< 1$}\\
	0	& \mbox{otherwise}\end{array}\right.
\end{equation}

\noindent where $P=P(\ec)$ is a fourth-order polynomial with $P(1)=0$
and $\alpha=2.71$ \citep{wid00}.

The NFW profile is infinite in extent and mass.  For model building it
is desireable to have a finite halo and so, following \citet{kin66},
we introduce an energy cutoff $E_h\equiv -\ec_h\equiv
-\epsilon_h\sigma_h^2$ and replace $f_{\rm NFW}$ with

\begin{equation}
\label{eq:fhalo}
f_{\rm halo}\left (\ec\right ) = \sigma_h^{-1}
a_h^{-2}{\cal F}_{\rm halo}\left (\ec/\sigma_h^2\right )
\end{equation}

\noindent  where

\begin{equation}\label{eq:nfw_df2}
{\cal F}_{\rm halo}\left (\epsilon\right ) = \left\{
	\begin{array}{ll} 
	{\cal F}_{\rm NFW}\left (\epsilon\right ) -
	{\cal F}_{\rm NFW}\left (\epsilon_h\right ) & 
	\mbox{for $\epsilon_h<\epsilon <1$}\\
	0 & \mbox{otherwise~.}\end{array}\right.
\end{equation}

\begin{figure}[p]
\begin{center}
\plotone{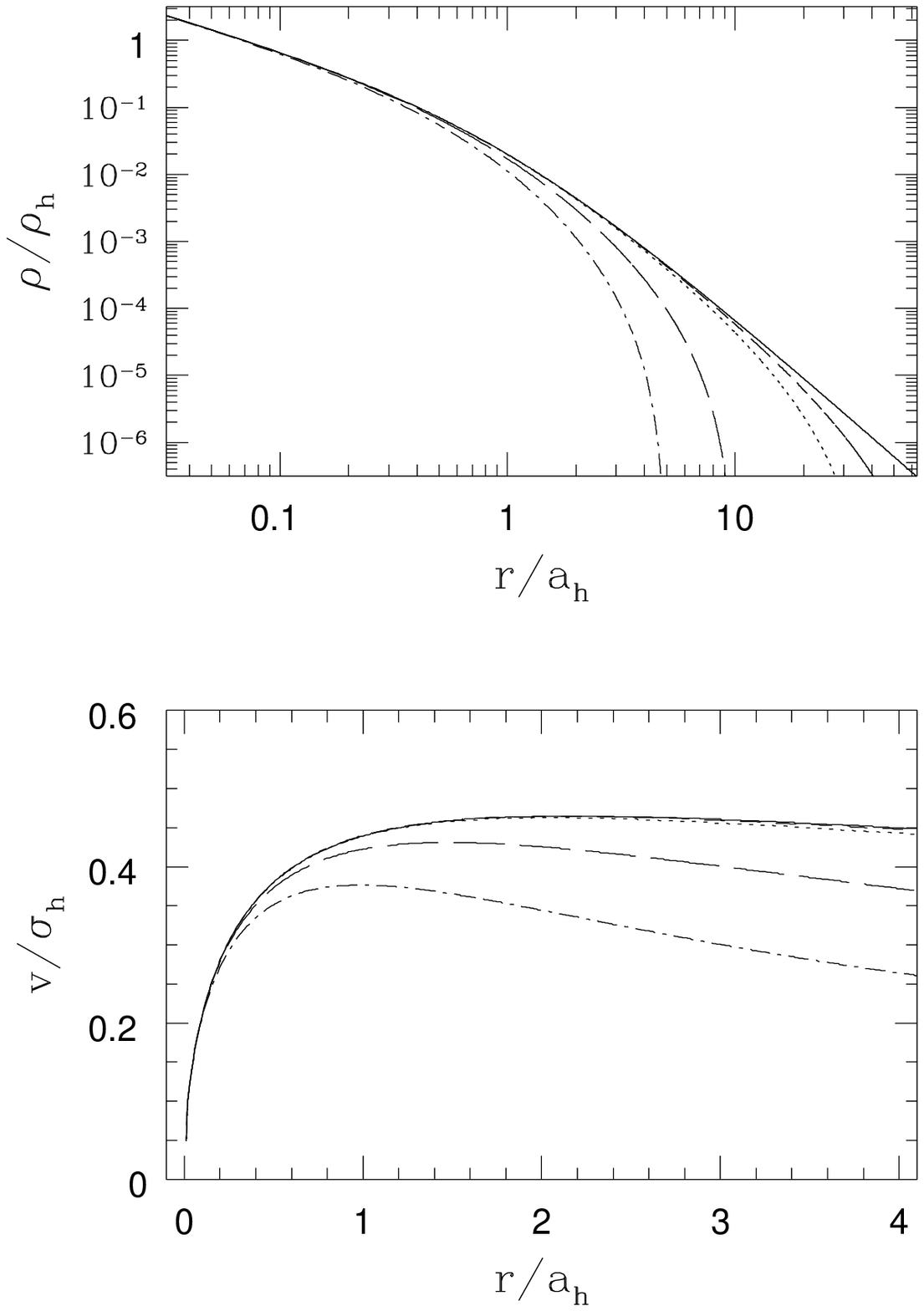} \caption{Density profiles (top) and rotation curves
(bottom) for halo models with $a_h=\sigma_h=1$ and values of
$\ec_h$ corresponding to tidal radii $r_t = 5, 10, 40$, and $80$.  Also
shown is the full NFW model (solid curve).\label{fig:isolatedhalo} }
\end{center}
\end{figure}

\noindent $\epsilon_h = 0$ corresponds to a full NFW profile while
$0<\epsilon_h<1$ yields a truncated profile.  Examples with
$a_h=\sigma_h=1$ and various values of $\epsilon_h$ are shown in
Figure~\ref{fig:isolatedhalo}.  Note that the truncation
radius can be varied independent of the characteristic density and
scale radius.  When building models a natural choice for the
truncation radius is the cosmologically-motivated virial radius,
$r_{\rm vir}$ (see below).  After the inner properties of a model
galaxy are set, a value of $\epsilon_h$ that gives $r_t \approx r_{\rm
vir}$ can be easily found.  In contrast, for the King and lowered
Evans models a change in the truncation radius results in a
significant change in the inner profile making model building more
cumbersome and less intuitive.

The DF as written above is symmetric under $J_z\to -J_z$ and therefore
generates a model with no net rotation.  By spitting the DF into parts
with positive and negative $J_z$ (${\cal F_\pm}$) and recombining them
with a suitable coefficient, one can generate a model with arbitrary
amounts of rotation.  Formally we write

\begin{equation}\label{eq:nfw_df3}
{\cal F}_{\rm halo} = \alpha_h {\cal F}_+ + \left (1-\alpha_h\right )
{\cal F_-}
\end{equation}

\noindent where $\alpha_h$ controls the amount of rotation
($\alpha_h=1/2$ implies no rotation).

\subsection{The bulge distribution function}

Bulges are commonly modelled as the de Vaucouleurs $r^{1/4}$ law
\citep{deV48} in projection and the Hernquist model is a simple
density-potential pair which closely mimics this behaviour
\citep{her90}.  We therefore model the bulge using a Hernquist DF
modified by a energy cutoff $E_b$ to truncate the profile in a 
fashion similar to what was done for the halo.

The standard Hernquist model has a density profile and potential given
by

\begin{equation}\label{eq:bulgeprofile}
\rho_{\rm H} = 
\frac{\rho_b}{\left (r/a_b\right )
	\left (1 + r/a_b\right )^3}
\end{equation}

\noindent and

\begin{equation}\label{eq:bulgepotential}
\Phi_{\rm H} = -\frac{\sigma_b^2}{1+r/a_b}
\end{equation}

\noindent where $a_b$, $\rho_b = \sigma_b^2/2\pi a_b^2$, and
$\sigma_b$ are the scale length, characteristic density, and
characteristic velocity of the bulge.  While the total mass is finite
the density distribution is infinite in extent with $\rho \propto
r^{-4}$ at large radii.

We modify the Hernquist DF by incorporating an energy cutoff:

\begin{equation}\label{eq:bulgeDF}
f_{\rm bulge}\left (\ec\right ) = \sigma_b^{-1}a_b^{-2}
{\cal F}_{\rm bulge}\left (\ec^{1/2}/\sigma_b\right )
\end{equation}

\noindent where

\begin{equation}\label{eq:bulgeDF2}
{\cal F}_{\rm bulge}\left (q\right ) = \left\{
	\begin{array}{ll} 
	{\cal F}_{\rm H}\left (q\right ) -
	{\cal F}_{\rm H}\left (q_b\right ) & 
	\mbox{for $q_b<q<1$}\\
	0 & \mbox{otherwise}~,\end{array}\right.
\end{equation}

\noindent $q_b \equiv \left (-E/\sigma_b^2\right )^{1/2}$, and ${\cal
F}{\rm H}$ is the infinite-extent Hernquist model DF,

\begin{equation}\label{eq:hernquist}
{\cal F}_{\rm H}(q)=\frac{1}{2^{7/2}\pi^3}
		\frac{1}{\left (1-q^2\right )^{5/2}}
		\left (3\sin^{-1}q+q\left (1-q^2\right )^{1/2}
		\left (1-2q^2\right )
		\left (8q^4 - 8q^2-3\right )\right )~.
\end{equation}

\noindent As with the halo, rotation is introduced through an additional
parameter $\alpha_b$.

It is straightforward to use other models for the bulge.  For example,
the potential and DF for the density profile $\rho\propto r^{-3/2}
\left (r+a\right )^{-5/2}$ are analytic and the associated surface
density profile provides a somewhat better match to the de Vaucouleurs
law than the surface density profile of the Hernquist model
\citep{deh93}.
	
\subsection{The disk distribution function}

The disk is assumed to be axisymmetric with space density $\rho_{\rm
disk} = \rho_{\rm disk}\left (R,z\right )$ and quasi-Maxwellian DF
taken directly from the KD models which, in turn, were based on the
2-D model by \citet{shu69} and extensions by \citet{bin87a}.  This DF
is a function of $E,~J_z,$\, and $E_z$, the latter being an
approximate third integral of motion that corresponds to the energy in
vertical oscillations.  An implicit assumption in the formulation of
this DF is that the velocity dispersions are small so that the
epicyclic approximation is valid in the treatment of disk star orbits.
This assumption limits the application of the models to cool
disks. The DF can be tuned to yield a space density of a desired
form.  As in KD, we assume that the surface density profile of the
disk is exponential in the radial direction with scale radius $R_d$
and truncation radius $R_{\rm out}$.  The vertical structure is given
approximately by ${\rm sech}^{2}\left (z/z_d\right )$ where $z_d$ is
the vertical scale height.  In all, five parameters define the space
density of the disk: $R_d$, $R_{\rm out}$, $z_d$, the mass $M_d$, and
a profile ``shape'' parameter $\delta R_{\rm out}$ which governs the
sharpness of the truncation.

The ``stars'' of the KD disks have nonzero velocity dispersions in the
radial, azimuthal, and vertical directions.  The dispersion in the
radial direction, $\sigma_R(R)$, is assumed to be exponential:
$\sigma_R^2(R) = \sigma_{R0}^2\exp(-R/R_\sigma)$.  For simplicity, we
set $R_\sigma = R_d$ in accord with observations by \citet{bot93}.
The dispersion in the azimuthal direction is related to $\sigma_R$
through the epicycle equations \citep{bin87} while the dispersion in
the vertical direction is set by the vertical potential gradient and
the vertical scale height.

In total, the DFs for the three components are characterized by
fifteen parameters which we collect for convenience in Table 1.

\subsection{DFs for the composite model}

An isolated model halo is constructed by solving
Poisson's equation self-consistently for the gravitational potential

\begin{equation}\label{eq:poisson}
\nabla^2\Phi_{\rm halo} = 4\pi\rho_{\rm halo}
\end{equation}

\noindent together with the integral relation between the DF
(Eq.\ref{eq:nfw_df2}), the density, and the relative potential
$\psi\equiv -\Phi$:

\begin{equation}\label{eq:density}
\rho_{\rm halo} = 2^{5/2}\pi\int_0^\psi d\ec 
\sqrt{\left (\ec-\psi\right )}
f_{\rm halo}\left (\ec\right )
\end{equation}

\noindent \citep{bin87}.  For a disk-bulge-halo model,
Eq.\,\ref{eq:poisson} is replaced by

\begin{equation}\label{eq:poisson2}
\nabla^2\Phi = 4\pi\rho
\end{equation}

\noindent where

\begin{equation}\label{eq:density2}
\rho = \int d^3v 
\left (f_{\rm disk}+f_{\rm bulge}+f_{\rm halo}\right )
\end{equation}

\noindent is the total density.

\noindent The DFs in this expression depend on the energy (as well as
other integrals of motion) which in turn depends on $\Phi$.  Since
$\Phi$ is the total gravitational potential, the density fields of the
different components are inter-related in a complicated way.  In fact
the components of a disk-bulge-halo model constructed from Eq.\,6
of KD and Eqs.\,\ref{eq:nfw_df2} and \ref{eq:bulgeDF} may bare little
resemblance to the corresponding isolated components, a situation that
is cumbersome for model building.

To alleviate this problem, we proceed as follows.  The DF for an
isolated halo is nonzero over the energy range $\ec\in
\{\ec_h,\,\sigma_h^2\}$.  Qualitatively, we may expect the energy
range of halo particles in the multi-component system to be extended
to $\ec\in \{\ec_h, \sigma_h^2+\sigma_b^2 + \sigma_d^2\}$ where
$\sigma_d^2\equiv M_d/R_d$ is the depth of the disk potential.  These
arguments suggest that we replace $f_{\rm halo} \left (\ec\right )$ of
Eq.\ref{eq:fhalo} by $f_{\rm halo} \left (\ec'_h\left (\ec\right )\right
)$ where $\ec'_h\left (\ec\right )$ is a function that maps the energy
of a particle in the composite system ($\ec$) to the energy of a
particle in the would-be isolated halo ($\ec'_h$).  In what follows,
we assume

\begin{equation}\label{eq:energymapping}
\ec = \ec'_h\left (1 + \frac{\sigma_b^2+\sigma_d^2}
{\sigma_h^4}\ec'_h\right )~.
\end{equation} 

\noindent By construction, at low binding energies $(\ec\to 0)$, 
$\ec'_h \simeq \ec$ whereas $\ec\to\sigma_h^2+\sigma_b^2+\sigma_d^2$
as $\ec'_h\to\sigma_h^2$.  For the bulge, we replace $f_{\rm
bulge}\left (\ec\right )$ with $f_{\rm bulge} \left (\ec'_b\left
(\ec\right )\right )$ with $\ec_b\left (\ec\right ) = \ec'_b +
\sigma_h^2 + \sigma_d^2$.

These mappings are by no means unique and, apart from the conditions
discussed above, are not motivated by any particular physical model.
Indeed other mappings can be shown to yield similar and equally
acceptable models.

\begin{figure}[p]
\begin{center}
\plotone{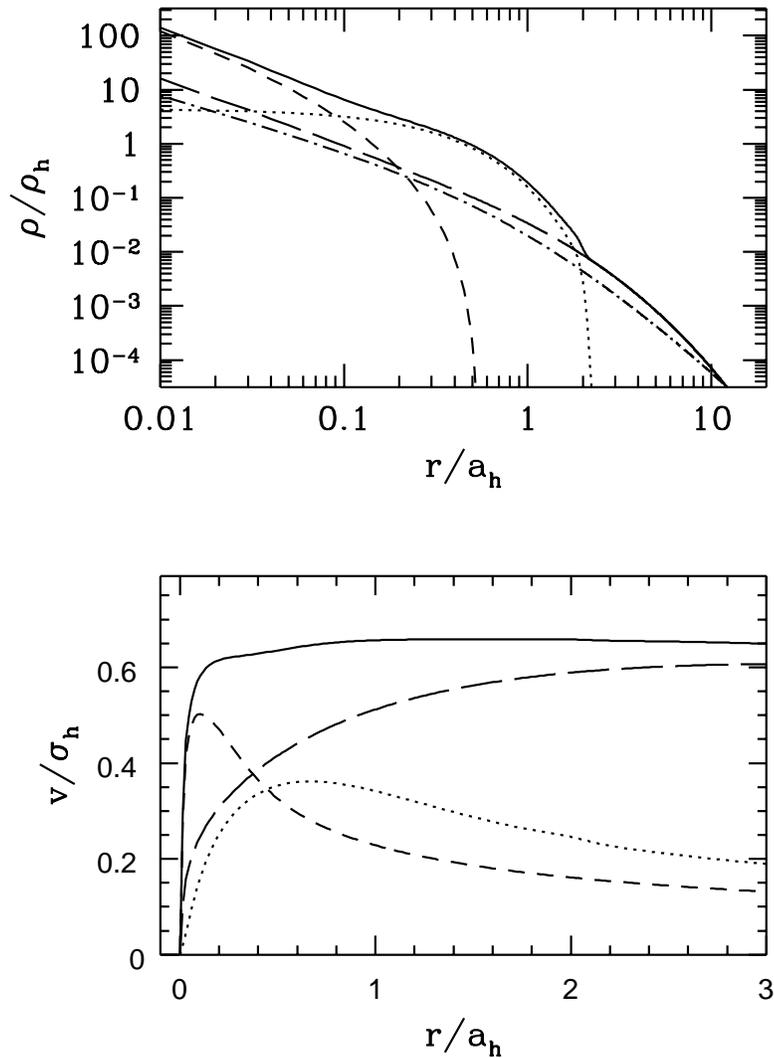} \figcaption{Density profile (top) and rotation curve
(bottom) for the disk-bulge-halo model described in the text.  Shown
are the contributions from the disk (dotted curve), bulge (dashed
curve), halo (long-dashed curve) as well as the total density profile
and rotation curve (solid curve).  Also shown is the halo model that
results if the same halo parameters are used and the disk and bulge
are not included (dot-dashed curve in the top
panel).\label{fig:figure2}}
\end{center}
\end{figure}

In Figure \ref{fig:figure2} we show an example of the density profile
and rotation curve for a typical disk-bulge-halo model.  (The model
parameters are given in Table 2.)  Also shown is the halo profile that
results assuming the same values of the $a_h$, $\sigma_h^2$ and
$\ec_h$ if the disk and bulge are not included.

\subsection{Comment on adiabatic compression}

It is important to stress that the ``isolated halo'' in Figure
\ref{fig:figure2} is not meant to represent the progenitor for the
halo in the final composite model.  Indeed, the structure of dark
halos in fully developed disk galaxies can be determined only through
detailed modeling of galaxy formation.  According to the standard
structure formation paradigm at early times baryons and dark matter
are well-mixed.  Disks form when the (collisional) baryons lose energy
(but not angular momentum) and settle to the bottom of the dark halo
potential well.  The halo in turn reponds to the change in the
gravitational potential and therefore a pristine NFW halo will
readjust to a new configuration.

Simulations of galaxy formation which include gasdynamics and star
formation are still relatively crude and computationally expensive.
An alternative is to treat the effect of baryon infall on dark
halos as an adiabatic process \citep{blu86, flo93}.  If changes in the
potential are slow compared with the orbital time of dark halo
particles, then for each particle, there is a set of quantites known
as adiabatic invariants which remain approximately constant even as
the orbit changes.  Knowledge of the adiabatic invariants allows one
to determine the final DF from the initial DF without having to model
the evolution of the system explicitly.

Under a rather restrictive set of assumptions, the adiabatic theorem
leads to the following simple relation:

\begin{equation}\label{eq:adiabatic}
\left (M_{\rm baryon}(r) + M_{\rm dm}\right )r = M_{\rm halo}(r_i)r_i
\end{equation}

\noindent \citep{you80, blu86, flo93}.  In this formula, $M_{\rm
baryon}(r)$ is the baryon mass distribution as a function of radius
(i.e., a spherically averaged representation of the disk and bulge),
$M_{\rm dm}$ is the final distribution of the dark halo, and $M_{\rm
halo}(r_i)$ is the initial distribution of the baryon-dark matter
proto-galaxy.  Eq.\ref{eq:adiabatic} allows one to go directly from the
cosmologically-motivated NFW halo to the final dark halo in a
fully-formed galaxy and has been used extensively in modeling disk
galaxies (See, for example, \citep{mmw98, kly02}.

In general, the dark halos predicted by Eq.\ref{eq:adiabatic} are more
centrally concentrated than the progenitors with a density profile
that is cuspier in the center (e.g., $r^{-1}$ cusp becomes an
$r^{-1.5}$ cusp as in Figure 7 of \citet{kly02}).  However
Eq.\ref{eq:adiabatic} is based on a number of suspect assumptions.  In
addition to spherical symmetry and adiabaticity, one must assume that
the dark halo particles are on circular orbits and therefore do not
cross as their orbits contract.  The particle orbits in simulated
halos are, if anything, radially biased.  (See \citet{moo04} who study
the importance of the distribution of orbits for the evolution of dark
halos.)

An indication of the uncertainty in the adiabatic prescription can be
seen in Figure 7 of \citet{kly02} where two halos from the same
progenitor are shown.  In one case, Eq.\ref{eq:adiabatic} is used to
determine the halo profile while in the other, baryons and dark matter
are allowed to exchange angular momentum.  The difference between the
two halos is significant with the former being denser in the central
regions by a factor of 30!

How do our models fit in with the adiabatic compression paradigm?  In
short, we side-step the issue by focusing on models of fully developed
disk galaxies.  One might use Eq.\ref{eq:adiabatic} to work backwards
from our final model and determine the progenitor (and thus compare
with the NFW profile) but given the caveats associated with this
equation we do not see this as a particularly fruitful endeavor.  We
do note that the difference between a model halo in isolation and one
incorporated into a composite system is consistent with the general
predictions of adiabatic compression theory --- the profile exhibits a
slightly cuspier inner region and is generally more concentrated ---
and so the naive use of the halo parameters as an indication of the
general structure of the progenitor halo seems a reasonable first pass
for making contact with cosmology.

\subsection{Models with central supermassive black holes}

The observation that most, if not all, disk galaxies harbor
supermassive black holes near their centers has led to considerable
interest in the structure of black hole-stellar systems as well as the
effect a black hole might have on the central cusp of a dark matter
halo.  \citet{tre94} have derived DFs for a variety of models with
central black holes whose density profiles have $r^{-\eta}$ central
cusps and $r^{-4}$ outer parts.  These so-called $\eta$-models include
the Hernquist profile ($\eta = 1$) and are therefore directly
applicable to the bulge DF in our system and easily generalized to the
NFW halo model.  The DFs of \citet{tre94} produce a density profile
that is independent of the black hole mass: For a given $\eta$-model,
the black hole alters the velocity distribution but not the space
distribution of the stars and dark matter particles in its vicinity.
We build black holes into our models under the same assumption.

Consider first the bulge-black hole system.  The effect of a black
hole of mass $M_{\rm BH}$ is to modify the relative potential

\begin{equation}\label{eq:psistar}
\psi^*(r) = \psi(r) + \frac{M_{\rm BH}}{r}
\end{equation}

\noindent where as in \citet{tre94} the superscript $*$ 
denotes properties of models that include a black hole.  While
the Hernquist DF is non-zero over the energy range $\ec\in
\{\ec_b,\,\sigma_b^2\}$, with a black hole, the energy range is
extended to $\ec^*\in \{\ec_b,\,\infty\}$.

In principle, disk-bulge-halo models with a central black hole can be
constructed by replacing $f_{\rm bulge}$ with a DF from \citet{tre94}
that is modified by a suitable energy cut-off.  The DFs for the disk
and halo would be similarly replaced.  However, the DFs in
\citet{tre94} are not analytic and would have to be recalculated for
each choice of $M_{\rm BH}$.  We therefore choose the following more
efficient, albeit {\it ad hoc} scheme: The potential and density
profile for a particular disk-bulge-halo model are calculated assuming
no black hole.  To incorporate a black hole, the potential is modified
according to Eq.\ref{eq:psistar} and a new DF is found which interpolates
between $f_{\rm bulge}$ of Eq.\,\ref{eq:bulgeDF} and the DF of the
appropriate $\eta$-model.

At large binding energies ($\ec^*\to\infty$) the Hernquist-black hole
DF has the asymptotic form \citep{tre94} $\lim_{\ec^*\rightarrow
\infty}f^* = f^*_\infty$ where 

\begin{equation}\label{eq:asymBH}
f^*_\infty\left (\ec^*\right ) =
	\frac{\sigma_b}{2^{5/2}\pi^3 M_{\rm BH}a_b}
	\left (\frac{\ec^*}{\sigma_b^2}\right )^{-1/2}~.
\end{equation}

\noindent On the other hand,

\begin{equation}\label{eq:asymHQ}
f_{\rm H}(\ec) \to \frac{3}{2^{5/2}\pi^2\sigma_b a_b^2}
		\left (1-\ec/\sigma_b^2\right )^{-5/2}
\end{equation}

\noindent for $\ec\to\sigma_b^2$.
From the asympototic form of the Hernquist potential, $\psi_{\rm H}
\simeq \sigma_b^2/\left (1-r/a_b\right )$, we have $r\simeq a_b\left
(1+\psi_{\rm H}/\sigma_b^2\right )$ suggesting

\begin{equation}\label{eq:bh_mapping}
\ec^* = \ec + \frac{M_{\rm BH}}{a_b}
\left (\frac{1}{1-\ec/\sigma_b^2}-\frac{1}{1-\ec_b/\sigma_b^2}\right )
\end{equation}

\noindent as a mapping from $\ec$ to $\ec^*$.  By construction,
$\ec^*(\ec_b)=\ec_b$.  We take the DF to be

\begin{equation}\label{eq:HernBH}
f^*(\ec^*) = {f_{b}(\ec)}
\left (1 + \frac{f_{b}(\ec)}{f_\infty^*\left (\ec^*\right )}
\right )^{-1}
\end{equation}

\noindent which smoothly interpolates between $f_b$ and Eq.\ref{eq:asymHQ}.
A similar procedure is carried out for the halo and disk.

\section{Models for the Milky Way and Andromeda Galaxies}

In this section we present models chosen to fit observational data for
the Milky Way and Andromeda galaxies.  The level of realism in our
models is dictated to a large extent by the assumptions upon which they
are based.  At first glance, the assumption of axisymmetry seems
rather restrictive since virtually all disk galaxies exhibit
non-axisymmetric phenomena such as bars and spiral arms.  However, our
models are subject to non-axisymmetric instabilities.  The program
adopted here is to find the best-fit axisymmetric model and determine,
through N-body simulations, if the model evolves to a state that more
closely matches the actual galaxy.  This approach has already been
applied to the Milky Way by \citet{sel85, sel93}, \citet{fux97}, and
\citet{val03}.

Another key assumption is that the model galaxies are comprised of
three components, a disk, bulge, and halo, whereas actual disk galaxies
also have stellar halos and globular clusters that are typically
spheroidal, more extended than the bulge, but (presumably) less
extended than the dark halo.  It would be straightforward to include
such systems in our models but since they contain relatively little
mass we do not do so here.  The inclusion of globular cluster
systems might be of interest for studying their evolution during 
galaxy mergers and interactions.

Along similar lines, the structural parameters of the galactic disks
(e.g., radial and vertical scale heights, velocity dispersion tensor)
depend on colour and metalicity \citep{deh98b} whereas our models
assume a single disk component.  Stellar disks may, in fact, be more
accurately represented as two component systems with a young thin disk
and an old thick disk.  Our models can be easily modified to include
two or more disk-like components, an improvement which may prove
relevant for detailed studies of bars and spiral structure.

Our models assume DFs for the bulge and halo that are functions of the
energy so that the velocity distribution in these components is
necessarily isotropic.  By contrast, the velocity distributions in
simulated halos are typically biased toward radial orbits with a
velocity anisotropy parameter $\beta\equiv 1 -
\overline{v_\theta^2}/\overline{v_r^2}\simeq 0.6$ \citep{vdb99}.
Spherical models with nonzero anisotropy parameters can be constructed
from functions of $E$ and $J$ where $J$ is the total angular momentum
\citep{osi79, mer85, bae02}.  However, $J$ is not an integral of
motion of a general axisymmetric.  Moreover, since disks rotate, we
can be fairly certain that halos do as well.  Thus, realistic halo
models require three or more integrals of motion.  The velocity
structure of dark matter particles will have an effect on the
dynamical interaction between the halo and stellar components (e.g.,
the decay of the bar pattern speed through dynamical friction) and so
it may be of interest to consider more general halo DFs.  We leave the
investigation of these subtle effects for future work.

Perhaps the most severe assumption is that our models include only
collisionless components whereas for some disk galaxies, a significant
fraction of the ``disk'' mass is locked up in neutral hydrogen gas.
Future versions of our models will include HI disks thereby
expanding the applications to models with gas as well as stars.

For each galaxy, we select a set of observational data which are
compared with pseudo-observations of the model galaxy to yield a
$\chi^2$-statistic.  Minimization of $\chi^2$ over the
multi-dimensional parameter space yields the desired best-fit model.
In addition, one can include non-observational constraints so as to select
models with certain characteristics (e.g., specific value for the
bulge-to-disk mass ratio, baryon fraction, or halo concentration
parameter).

In addition to the 15 parameters of Table 1 observer-dependent
parameters such as the inclination angle for external galaxies or the
galactocentric radius of the Sun for Milky Way must also be specified.
In addition, if photometric data is used, then the mass-to-light
ratios of the disk and bulge are also required.  Depending on the type
of observations, some of the parameters may be fixed during the
minimization process.

Following \citet{wid03} we employ the downhill simplex algorithm (see,
for example, \citet{pre86}) for the minimization of $\chi^2$.  A
simplex is a geometrical figure in $N$ dimensions where $N$ is the
number of parameters that defines the model.  During execution of the
algorithm, the simplex moves through parameter space seeking out the
minimum of $\chi^2$.  As it does so, the simplex changes shape thus
enabling it to move through complicated regions of parameter space.
The downhill simplex method has a number of advantages over
minimization procedures based on gradients of $\chi^2$ (e.g., the
method of steepest descent; see \citet{pre86} and references therein).
In particular, the method appears to be less susceptable to false
minima though restarts are always executed to guard against this
possibility.

\subsection{The Milky Way}

Numerous authors have attempted to model the Milky Way with early
important studies by \citet{bah80} and \citet{cal81}.  \citet{bah80}
fit star count data together with local values for the scale heights
and luminosity functions to parameterize global models of the disk
and bulge.  \citet{cal81} considered dynamical constraints (e.g., the
Oort constants, rotation curve) in constructing three-component mass
models.  More recently, \citet{deh98a} improved and updated these
models by augmenting rotation curve data with dynamical constraints on
the vertical structure of the Galaxy in the solar neighborhood.

\subsubsection{Observational constraints}

The models developed here for the Milky Way are assembled in the same
spirit of early mass models but have the advantage that the end
result is a fully realized DF for the stars and dark matter of the
Galactic system.  Seven observational data sets are used to constrain
the Milky Way models.  Five of these data sets, the inner and outer
rotation curves, the Oort constants, the vertical force in the solar
neighborhood, and the total mass at large radii are taken directly
from \citet{deh98a} and references therein.  We also use measurements
of the bulge dispersion at a projected distance of $200\,{\rm pc}$
from the galactic center (the peak of the dispersion profile) from the
compilation of data by \citet{tre02}.  Finally, we incorporate
estimates of the local velocity ellipsoid from \citet{bin98}.

\begin{itemize}

\item{\it Inner rotation curve}

Observations of HI emission provide a direct measure of the Galactic
rotation curve.  Inside the solar circle these observations are
usually presented in terms of the so-called terminal velocity, $v_{\rm
term}$, the peak velocity along a given line-of-sight at galactic
coordinates $b=0$ and $|l|<\pi/2$.  Assuming the Galaxy is
axisymmetric and the ISM rotates on circular orbits, the HI emission
corresponding to $v_{\rm term}$ originates from the galactocentric
radius $R=R_0\sin{l}$.  Relative to the local standard of rest,
we have

\begin{equation}\label{eq:vterm}
v_{\rm term} = v_c(R)-v_c\left (R_0\right )\sin{l}
\end{equation}

\noindent where $v_c$ is the circular speed (see, for example,
\citet{bin98}).  Following \citet{deh98a} we use data from
\citet{mal95} restricted to the range $\sin{l}\ge 0.3$ so as to avoid
distortions from the bar.  In particular, we use values of $v_{\rm
term}$ from Figure 7 of her paper at four representative values of
$|\sin{l}|$ averaging the results from the first and fourth quadrants.

\item {\it Outer rotation curve}

The radial velocity of an object at Galactic coordinates $\left
(l,b\right )$ relative to the local standard of rest, $v_{\rm lsr}$ is
related to the circular rotation curve through the equation

\begin{equation}\label{eq:outerRC}
v_{\rm lsr}=\left (
\frac{R_0}{R}v_c\left (R\right )-
v_c\left (R_0\right )\right )\cos{b}\sin{l}
\end{equation}

\noindent where $R=\left (d^2\cos^2{b}+R_0^2-2R_0d\cos{b}\sin{l}\right
)^{1/2}$ and $d$ is the distance to the object.

In general, the data consists of a set of measurements ($v_{{\rm
lsr},i},~ d_i$) which is to be compared with the model estimate $W(R)$
and $d(R)$ where $R$ may be regared as a free parameter and $W(R)\equiv
\left (R_0/R\right )v_c(R)-v_c(R_0)\equiv v_{lsr}/\cos{b}\sin{l}$.
For each data point, we adjust $R$ so as to minimize

\begin{equation}\label{eq:chi_i}
\chi^2_i = \left (\frac{W(R)-W_i}{\Delta W_i}\right )^2+
\left (\frac{d(R)-d_i}{\Delta d_i}\right )^2
\end{equation}

\noindent where $W_i\equiv v_{{\rm lsr},i}/\cos{b}\sin{l}$.  The
$\chi^2$ for this data set is the average of the individual
$\chi_i^2$'s.

In what follows we use data from \citet{bra93} with the same
restrictions as in \citet{deh98a} (i.e., $l\le 155^\circ$ or $l\ge
205^\circ$, $d>1\,{\rm kpc}$, and $W<0$).

\item {\it Vertical force above the disk}

\citet{kui91} used K dwarf stars as tracers of the gravitational
potential above the galactic plane thereby placing a constraint on the
total mass surface density in the solar neighborhood.  They found

\begin{equation}\label{eq:kuigil1}
\frac{|K_z\left (1.1\,{\rm kpc}\right )|}{2\pi G}
= 71\pm 6\,M_\odot {\rm pc}^{-2}
\end{equation}

\noindent independent of the relative contributions of the disk and
halo.  Only by including additional constraints on the local circular
speed, galactocentric distance of the Sun, and Oort constants can one
ferret out the separate contributions of the disk and halo to the
local surface density.  Doing so, \citet{kui91} found

\begin{equation}\label{eq:kuigil2}
\Sigma_{\rm disk}
= 48\pm 9\,M_\odot {\rm pc}^{-2}
\end{equation}

\noindent which is in excellent agreement with estimates of known
matter in the solar neighborhood.  Since we include constraints on the
Oort constants and rotation curve separately, we use Eq.\ref{eq:kuigil1}
as a constraint on our models and Eq.\ref{eq:kuigil2} as a consistency
check.

\item {\it Oort constants}

The Oort constants

\begin{equation}\label{eq:oort_a}
A \equiv \frac{1}{2}
	\left (\frac{v_c}{R}-\frac{\partial v_c}{\partial R}\right )
\end{equation}

\noindent and

\begin{equation}\label{eq:oort_b}
B \equiv -\frac{1}{2}
	\left (\frac{v_c}{R}+\frac{\partial v_c}{\partial R}\right )
\end{equation}

\noindent measure, respectively, the shear and vorticity in the Galactic
disk.  Following \citet{deh98a}, who review the published measurements,
we adopt the constraints

\begin{equation}\label{eq:oort_contraints}
A = 14.5\pm 1.5\,{\rm km\,s^{-1}\,kpc^{-1}}~~~~~~~
B = 12.5\pm 2\,{\rm km\,s^{-1}\,kpc^{-1}}~.
\end{equation}

\item {\it Local velocity ellipsoid}

The kinematics of stars in the solar neighborhood provides important
constraints on the structure of the Milky Way.  The observation that
$\overline{v_R^2}\ne \overline{v_z^2}$ already tells us that the disk
DF cannot take the form $f=f\left(E,L_z\right )$ and necessarily
involves a third integral of motion \citep{bin87}.  Since the KD disks
are built from three-integral DFs it is possible to model anisotropic
velocity dispersion in galactic disks.  There are two important
caveats.  The first, already mentioned in the introduction to this
section, is that the models assume a single disk component so that the
velocity dispersions in the radial, azimuthal, and vertical directions
are single-valued functions of cylindrical radius.  In point of fact,
the shape of the velocity ellipsoid depends of colour (see, for
example, \citet{deh98b}).  Furthermore, the velocity ellipsoid is
rotated about the z-axis so that the principle axes do not coincide
with the $\hat{R}$ and $\hat{\phi}$ directions.  The ``vertex
deviation'' varies from $0^\circ-30^\circ$ depending on $B-V$ colour.

These limitations can be overcome without too much difficulty (e.g.,
by including more than one disk component and by generalizing the disk
DF to allow for vertex deviation of the velocity ellipsoid).  We leave
these improvements for future work and consider a single component
disk.  For constraints on the velocity ellipsoid, we use values from
Table 10.4 of \citet{bin98} which were derived from \citet{edv93}.
\citet{bin98} give values for the thin and thick disks and we use a
mass-weighted average assuming a 14:1 ratio between these two
components \citep{deh98a} with 15\% $1\sigma$ errorbars.

\item {\it Bulge dispersion}

Observations of the LOSVD in the direction of the bulge provide
important constraints on the bulge parameters (and to a lesser extent,
the parameters of the other components).  \citet{tre02} have compiled
measurements of the LOSVD between $0.085$ and $1300\,{\rm pc}$.  The
dispersion profile shows a minimum of $\sim 55\,{\rm km s^{-1}}$ at
$r\simeq 5\,{\rm pc}$ and a maximum of $130\,{\rm km s^{-1}}$ at
$r\simeq 200\,{\rm pc}$ (see, also, \citet{ken92}).  The rise of the
dispersion profile inside $r\simeq 5\,{\rm pc}$ is presumably due to
the central black hole while the detailed shape of the dispersion
profile at larger radii may be affected by the barlike shape of the
bulge.  With this in mind, we average values of the line-of-sight
dispersion near the peak to arrive at the single constraint
$\sigma_{los}(R=210\,{\rm pc})=136\pm 12\,{\rm km\,s^{-1}}$.

\item {\it Mass at large radii}

The system of satellite galaxies that surround the Milky Way, the
Magellanic Stream, and the high-velocity tail of the local stellar
velocity distribution provide constraints on the large-scale mass
distribution of the Galactic halo.  Following \citet{deh98a}, who base
their arguments on work by \citet{koc96} and \citet{lin95}, we adopt

\begin{equation}\label{eq:mass100} 
M\left (r<100\,{\rm kpc}\right ) = \left (7\pm 2.5\right
)\times 10^{11} \,M_\odot 
\end{equation}

\noindent as a constraint on the mass distribution at large radii.

\end{itemize}

\subsubsection{Search strategy}

A $\chi^2$ statistic is calculated by comparing each of the seven data
sets described above with pseudo-observations of the model.  The
pseudo-observations are designed to match closely the actual
observations.  For example, the LOSVD in the bulge region is found by
calculating the velocity dispersion along a given line of sight of
bulge ``particles'' chosen from the DF.  In principle, one can add
additional layers of realism to the pseudo-observations such as
aperature smoothing for LOSVD measurements.

The results are averaged in quadrature to yield a composite
$\chi^2$-statistic.  In addition to the 15 parameters in Table 1 we
must also specify the galactocentric radius of the Sun, $R_0$.  We fix
$R_{\rm out}$ and $\delta R_{\rm out}$ to $30\,{\rm kpc}$ and $1\,{\rm
kpc}$ respectively.  Since the surface density of the disk falls
exponentially, varying these values will not affect the model fit.  In
addition, the rotation parameters of the halo and bulge are fixed so
that neither component has net angular momentum.  (In principle,
incorporating more detailed observations of the bulge would allow us
to fit the bulge rotation curve.)  We pin the scale length of the
radial dispersion profile to half that of the disk scale length (i.e.,
$R_\sigma= R_d$ so that the $\sigma_R^2$ and the surface density have
the same exponential decay constant).  We also fix $\epsilon_h=0.2$
which gives a tidal radius larger than $100\,{\rm kpc}$.  Finally, we
run the simplex algorithm with $R_0$ fixed to different values.  In
summary, each implementation of the simplex algorithm is run with nine
free parameters.

\subsubsection{Results}

\begin{figure}[p]
\begin{center}
\plotone{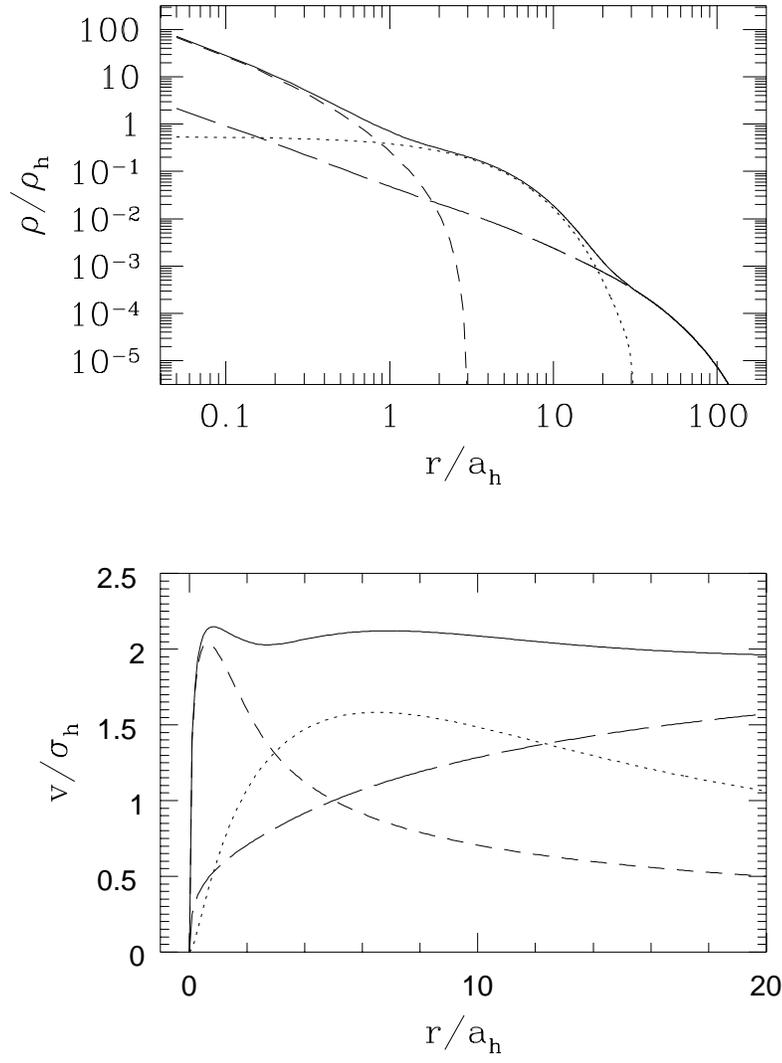}
\figcaption{Density profile and rotation curve for model MWa.
Line types are the same as in Figure 2.\label{fig:MWa-rho-vc}}
\end{center}
\end{figure}

\begin{figure}[p]
\begin{center}
\plotone{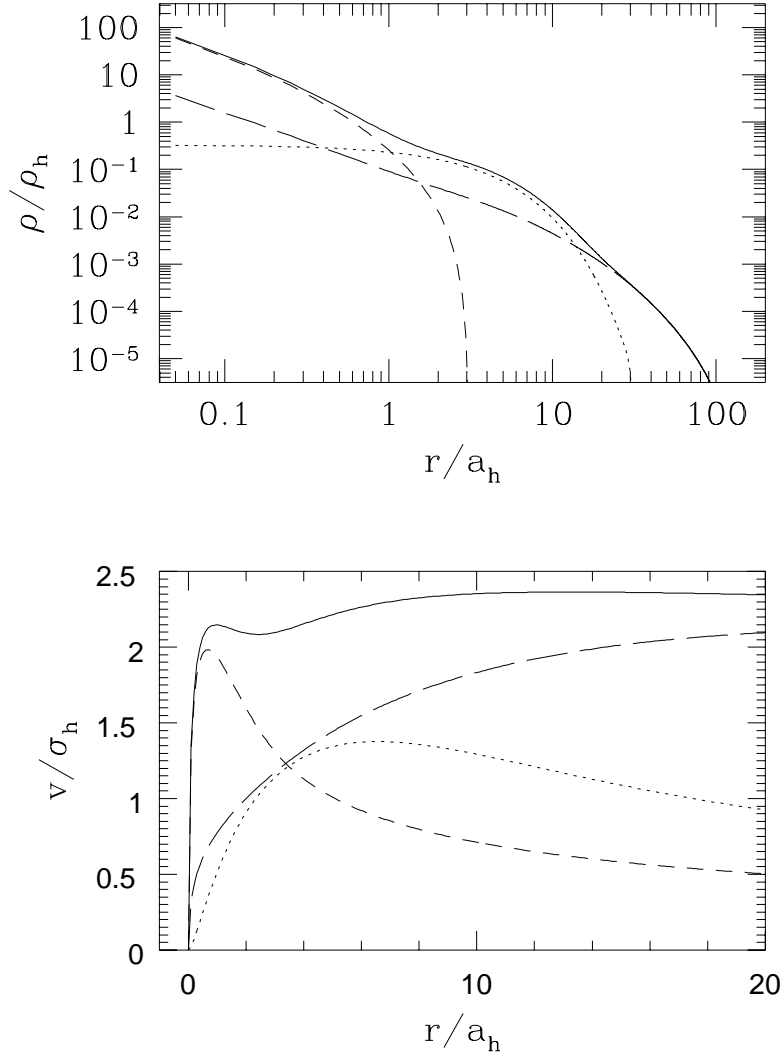} \figcaption{Density profile and rotation
curve for model MWb.\label{fig:MWb-rho-vc}}
\end{center}
\end{figure}

The parameter set for the best-fit $R_0=8\,{\rm kpc}$ model (MWa) is
given in Table 2.  Also given is the parameter set for a Milky Way
model with a less massive disk that also has an acceptable $\chi^2$
(MWb).  The density profiles and rotation curves for the two models
are shown in Figures \ref{fig:MWa-rho-vc} and \ref{fig:MWb-rho-vc}.
We see that the disk dominates the rotation curve in model MWa for
$3\,{\rm kpc} \la R\la 12\,{\rm kpc}$ whereas the disk never dominates
the rotation curve of model MWb.  We will return to this point below.

\begin{figure}[p]
\begin{center}
\plotone{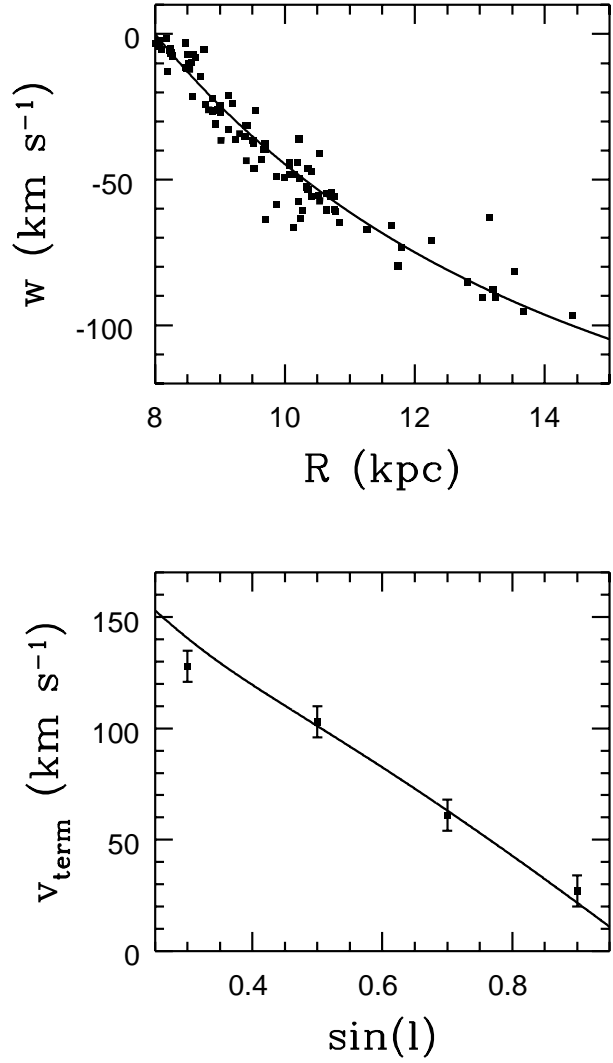} \figcaption{Results from the best-fit
Milky Way model compared with observations for $W$ (top) and
$v_{\rm term}$ (bottom) as defined in the text.\label{fig:MW-vterm-vouter}}
\end{center}
\end{figure}

A comparison of observations of $v_{\rm term}$ and $W(R)$ with model
predictions for MWa are shown in Figure \ref{fig:MW-vterm-vouter}
while a comparison of observed and predicted quantities for the other
observables are given in Table 3.  We see that both models provide an
excellent fit to the observations.

\begin{figure}[p]
\begin{center}
\plotone{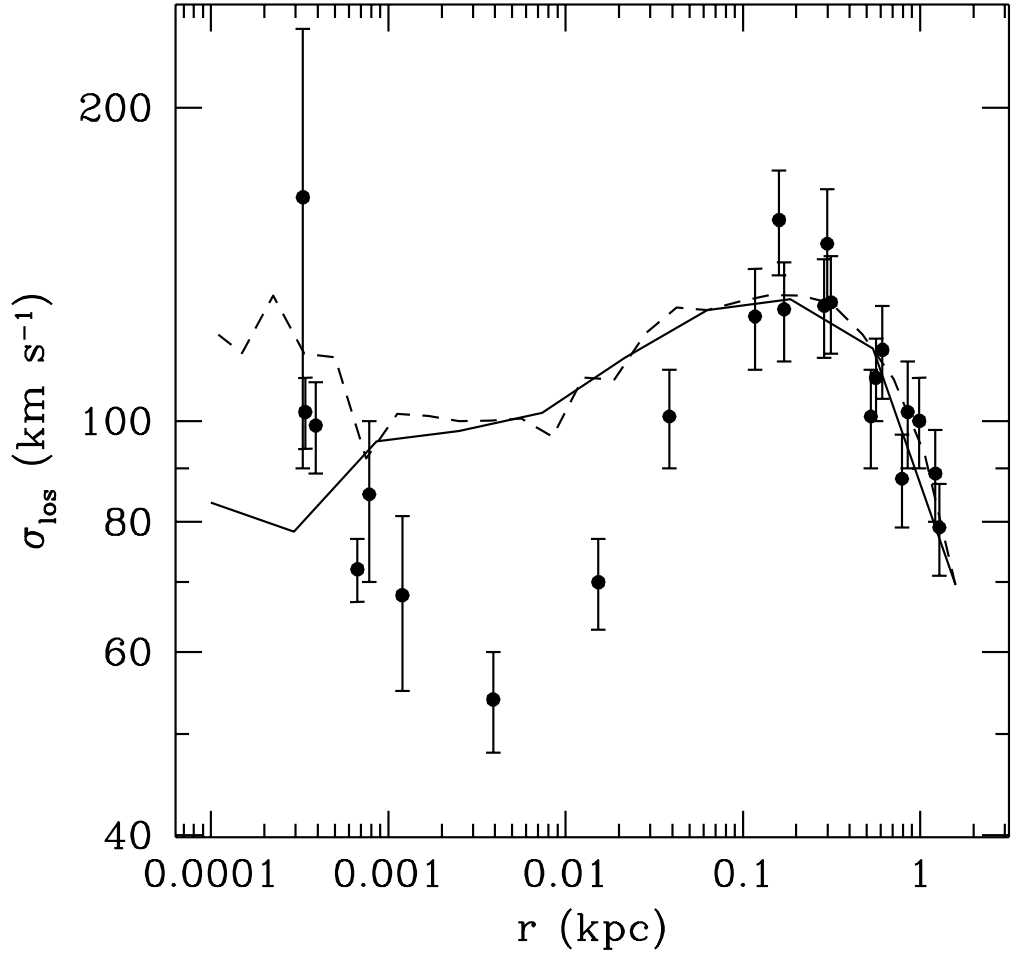}
\figcaption{Line of sight dispersion profile for the bulge
region.  Data are from a compilation by \citet{tre02}
of published measurements.\label{fig:bulgedisp}}
\end{center}
\end{figure}

In Figure \ref{fig:bulgedisp} we compare the line-of-sight velocity
dispersion profile for model MWa with data compiled by \citet{tre02}.
Recall that to find the model, a single constraint at $210\,{\rm pc}$
was used.  We see that the agreement between model and observations is
excellent for $r>100\,{\rm pc}$.  However, the model dispersion
profile at smaller radii is too flat in comparison with the data.  The
slow rise in the dispersion profile appears to be a feature our models
have in common with the $\eta$-models of \citet{tre94} and may point to
the necessity for a more complicated DF (e.g., one that depends on 
two or more integrals of motion).

Figure \ref{fig:bulgedisp} also includes the dispersion profile
obtained for a model that includes a central black hole of mass
$3.6\times 10^6\,M_\odot$.  This value is roughly twice that found by
\citet{tre02}.  The agreement between model and observations is poor
between $1\,{\rm pc}$ and $100\,{\rm pc}$ perhaps for the
aforementioned reasons.  However, the model profile does exhibit a
minimum at $r\simeq 10\,{\rm pc}$, a feature that is generic to
galaxies that are believed to harbor central supermassive objects.

\begin{figure}[p]
\begin{center}
\plotone{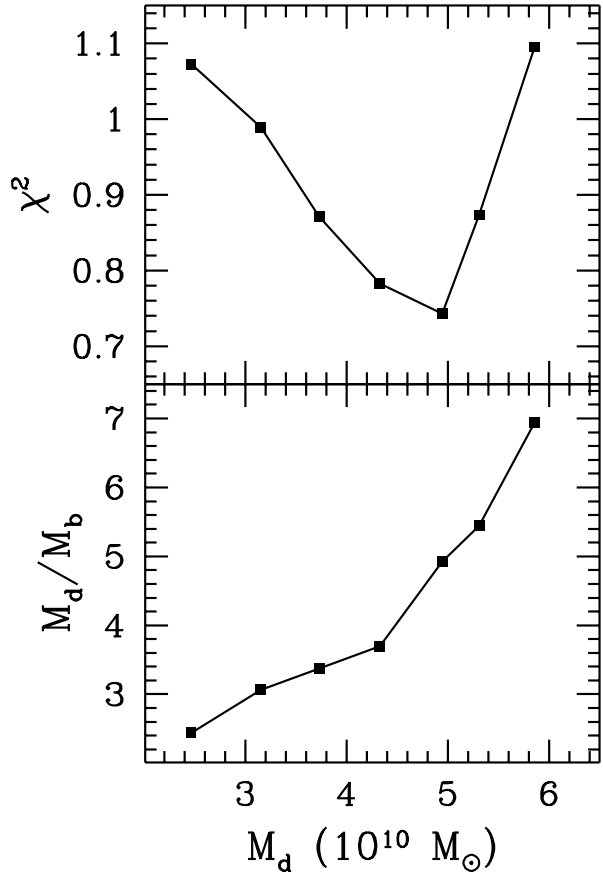} \figcaption{$\chi^2$-statistic as a
function of $M_d$ (top).  The best-fit model is found with $M_d$ fit.
For each case, we show the bulge mass $M_b$ as a function of
$M_d$.\label{fig:chi-vs-mdisk}}
\end{center}
\end{figure}

\begin{figure}[p]
\begin{center}
\plotone{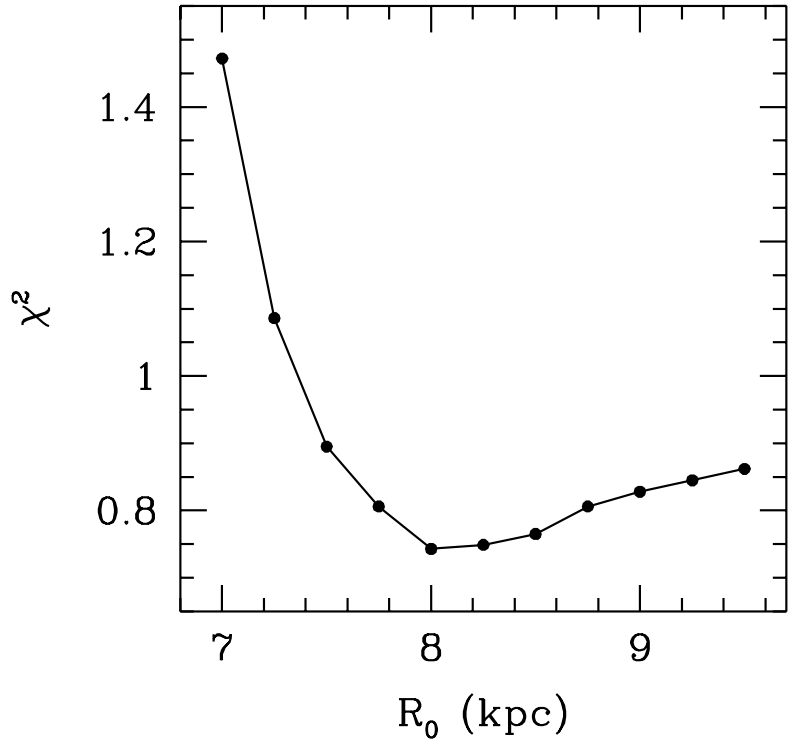}
\figcaption{$\chi^2$-statistic as a function of our galactocentric
radius, $R_0$.\label{fig:chi-vs-R0}}
\end{center}
\end{figure}

How tightly do the data constrain the model?  The answer in part is
given in Figures \ref{fig:chi-vs-mdisk} and \ref{fig:chi-vs-R0}.  For
Figure \ref{fig:chi-vs-mdisk}, $M_d$ is fixed to different values
while the remaining eight free parameters are allowed to float.  The
result is $\chi^2$ as aa function of $M_d$ as shown in the upper
panel.  The derived disk-to-bulge mass ratio, again as a funtion of
$M_d$ is shown in the lower panel.  Evidently, the disk mass can vary
by nearly a factor of two and still yield an acceptable fit to the
data.

A similar analysis for $R_0$ is shown in Figure \ref{fig:chi-vs-R0}.
Here, the minimum in $\chi^2(R_0)$ fairly flat especially toward
larger values ot $R_0$.  The preferred value of $R_0\simeq 8\,{\rm
kpc}$ agrees with well-known estimates.

For both MWa and MWb, $R_d/R_0\simeq 0.36$ which is approximately
halfway between the two NFW models considered by \citet{deh98a} (their
models 2d and 4d).  Table 4 summarizes values for a number of other
derived quantities for our models.  Column 2 gives the bulge mass.
The values for these two models are in excellent agreement with
estimates of $M_b$ based on COBE/DIRBE measurements \citep{dwe95} and
gravitational microlensing observations \citep{bis97}.

Column 3 gives the Toomre Q parameter at $2.5R_d$.  Both
models have $Q>1$ indicated that they are stable against local
perturbations \citep{bin87}.  Column 4 gives the local surface density
of the disk.  Our models evidently bracket the estimates from
\citet{kui91}.  

Columns 5 and 6 give the total mass and tidal radius for the models.
Recall that the tidal radius is controlled by the parameter
$\epsilon_h$.  We have tuned this parameter so that $r_t$ is roughly
equal to the virial radius $R_{\rm vir}$ of the Galaxy as predected by
the standard cosmological model of structure formation.  In this
scenario, $R_{\rm vir}$ and the mass interior to this radius, $M_{\rm
vir}$, are related through the equation

\begin{equation}\label{eq:mvir}
M_{\rm vir} = \frac{4\pi}{3}\Delta_{\rm vir}\,
\overline{\rho}\,R_{\rm vir}^3
\end{equation}

\noindent where $\overline{\rho}$ is the mean density of the Universe
and $\Delta_{\rm vir}$ parametrizes the average overdensity of the
halo. (See \citet{bul01} and references therein).  For the currently
favored $\Lambda$CDM cosmology ($\Omega_\Lambda=0.3$, $\Omega_m=0.7$)
one finds $\Delta_{\rm vir}=337$.  For both of our Milky Way models, 
$r_t\simeq 240\,{\rm kpc}$.

Based on these values of $r_t$, we can estimate the halo concentration
parameter $c_{\rm vir}\simeq r_t/a_h$ (Column 7).  Model MWa
compensates for the heavy disk by choosing a larger halo scale length
and for this reason, its concentration parameter is smaller than that
for MWb.  Both models have concentration parameters larger than the
mean value derived from cosmological simulations but within the
1$\sigma$ errorbars \citep{bul01}.

\subsubsection{Local dark matter density and the terrestrial
dark matter detectors}

Dark matter detection experiments rely on halo models to develop
search strategies and to analyse and interpret experimental data.
Terrestrial dark matter detectors are sensitive to the local density
and velocity-space distribution of dark matter particles.  For these
experiments, researchers have settled on a standard reference model,
namely Maxwellian velocities with an rms speed of $270\,{\rm
km\,s^{-1}}$ and a local density of $\rho(R_0)=0.0079\,M_\odot\,{\rm
pc}^{-3}\simeq 0.3\left (GeV/c^2\right )\,{\rm cm^{-3}}$.  This
reference model is useful for comparing the
sensitivities of different experiments as well as making contact with
predictions from theoretical particle physics.

\begin{figure}[p]
\begin{center}
\plotone{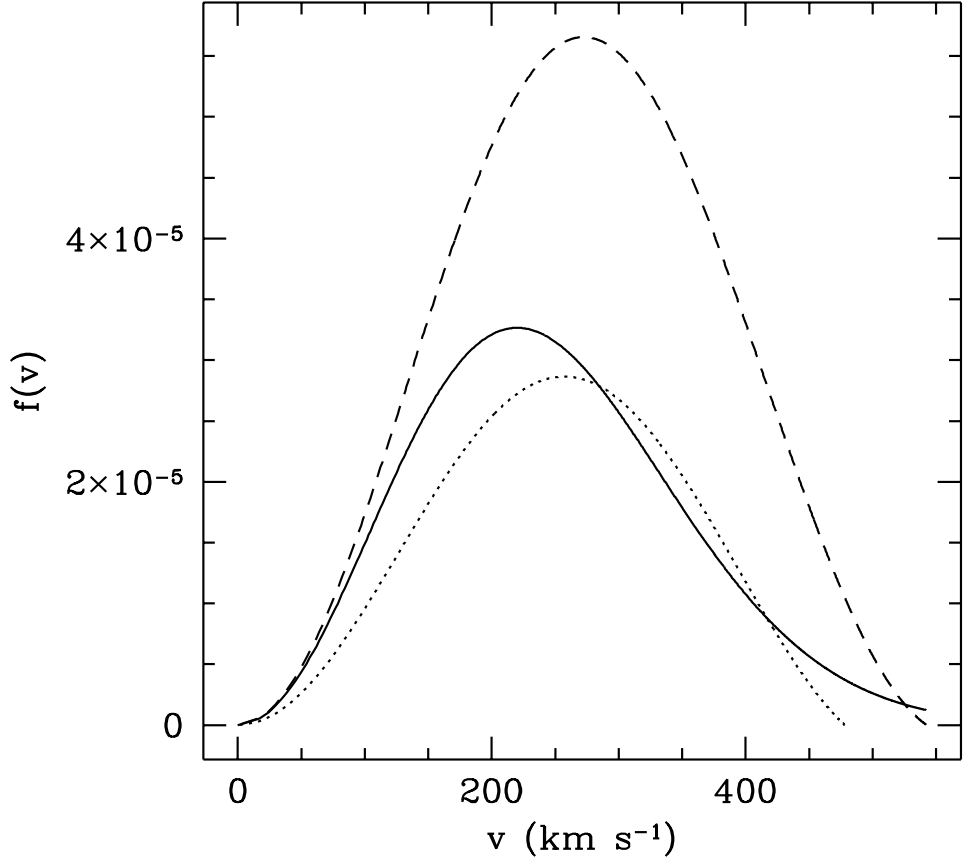} \figcaption{Speed distribution of halo
particles at the position of the solar system for model M31a (dotted
curve), M31b (dashed curve), and the standard model used by
experimentalists (solid curve).\label{fig:dmdetection}}
\end{center}
\end{figure}

Our models allow one to study deviations from the standard model while
ensuring that observational and dynamical constraints are satisfied.
In Figure \ref{fig:dmdetection} we compare the local speed
distribution for halo particles in models MWa and MWb with that of the
standard reference model.  The corresponding density is given in
Column 8 of Table 4.  We see that the heavy disk model (MWa) matches
up quite well with the standard reference model while the light disk
model has a local dark matter density that is a factor of two higher.
As discussed above, MWa is unstable to bar formation suggesting
a higher value for the local dark matter.

Numerous authors have considered variations on the standard Galactic
model such as bulk rotation of the halo, velocity space anisotropy,
triaxiality, tidal streams, and small-scale clumpiness.  Our models
provide a starting point for further investigations along these lines.
As described above, rotation may be added to the halo while
velocity-space anisotropy and triaxiality require a non-trivial
modification of the DF.  In addition, our models are well-suited to
numerical studies of the tidal disruption of subclumps in the dark
halo.

\subsection{Models for M31}

We now take up the challenge of modeling M31 which provides a case
study of an external galaxy.  The surface brightness profile, the
circular rotation curve, and the bulge velocity and dispersion
profiles are combined to yield observation-driven models.

Recently, \citet{wid03} combined observations of the types described
above to identify a suite of M31 models drawn from the original KD
set.  As one might expect, suitable models were found over a wide
range in values for the disk mass-to-light ratio though models with
particularly large values were found to be unstable to the formation
of a strong bar and could therefore be ruled out.

Our disk-bulge-halo DFs offer the possibility for an improved model of
M31.  The Hernquist bulge is favored by observations while the NFW
halo is favored by cosmological simulations of structure formation.
Morover, with the new models, the truncation radius of the NFW halo
can be varied independent of the inner density profiles of the three
components (Figure \ref{fig:isolatedhalo}).  Thus, one can tune the
truncation radius to correspond to the virial radius as predicted by
cosmology or alternatively, constrain the outer halo using
observations of the M31 satellite system.

\subsubsection{Observations}

Following \citet{ken89} and \citet{wid03} we utilize measurements of
Andromeda's surface brightness profile, rotation curve, and bulge
velocity profiles.  We use the global R-band surface brightness
profile from \citet{wal87} which was obtained by averaging the light
distribution over elliptical rings assuming an inclination of
$77^\circ$. 

The global surface brightness profile represents a small subset of the
available photometric data.  In principle a more sophisticated
bulge-disk decomposition, where the thickness of the disk and
inclination angle enter as free parameters, could be performed using
two-dimensional surface brightness maps.  However, apparent deviations
of the galaxy from axisymmetry suggest that our models cannot
adequately reflect this level of detail.  For example, the position
angle of the major axes of elliptical isophotes varies with
galactocentric radius suggesting that the bulge is barlike and
triaxial.  Triaxiality can be introduced in a controlled way, for
example by ``adiabatically molding'' the model into the desired shape
\citep{hol01}.  Alternatively one can evolve, via N-body methods, a
model galaxy that has a weak bar instability to see if some
intermediate state is consistent with the observations (e.g., twisting
isophotes).  Such an exercise might further constrain the models and
help break the mass-to-light degeneracy.

For the rotation curve we combine measurements by \citet{ken89} and
\citet{bra91} using the same kernel smoothing as in \citet{wid03}.
The composite rotation curve extends from $2-25\kpc$ in galactocentric
radii.  Though the measurements in \citet{bra91} extend to $30\kpc$,
we ignore data beyond $20\kpc$ since for this region of the galaxy,
measurements were made along a single spiral arm on one side of the
galaxy.  Stellar rotation and velocity dispersion results from
\citet{mce83} are used to constrain the dynamics of the inner $2\kpc$
of the galaxy.  \citet{wid03} attempted to fit the dispersion profiles
along the major and minor axes as well as the bulge rotation profile
between $300$ and $2000\,{\rm pc}$.  Acceptable fits were found though
in general, the models had a difficult time simultaneously reproducing
a rising rotation curve and falling dispersion profile.  One possible
explanation is that the ``bulge'' data is contaminated by the disk (a
rapidly rotating, dynamically cold system).  Here we use data
between $300$ and $1000\,{\rm pc}$.

The composite $\chi^2$ statistic is given by

\begin{equation}\label{eq:chi2}
\chi^2 = \frac{1}{3^{1/2}}\left (\chi^2_{\rm SPB} +
\chi^2_{\rm RC} +
\chi^2_{\rm B}\right )^{1/2}
\end{equation}

\noindent where $\chi^2_{\rm SPB}$, $\chi^2_{\rm RC}$, and
$\chi^2_{\rm B}$ for the surface brightness profile, rotation curve,
and bulge bulk velocity and dispersion profiles.

\subsubsection{Search strategy}

Two of the seven parameters that describe the disk DF, the mass and
radial scale length, are allowed to vary in the parameter-search
algorithm.  The vertical scale length, truncation radius, and
truncation shape parameter are fixed at the values $0.3$, $30$, and
$1\kpc$ respectively.  Since our data do not depend on the dispersion
of disk stars, the parameters $\sigma_{R0}$ and $R_\sigma$ are not
required for the fitting algorithm though they are required for
generating an N-body realization.  The halo parameters $\sigma_h$ and
$a_h$ are allowed to vary while the truncation parameter $\epsilon_h$
is fixed to a value large enough so that the truncation radius is well
outside the visible part of the galaxy.  This parameter is adjusted
after a suitable model for the visible part of the galaxy is found.
Finally, the four parameters that describe the bulge, $a_b$,
$\sigma_b$, $f_b$, and $\alpha_b$ are allowed to vary.

\subsubsection{Results}

\begin{figure}[p]
\begin{center}
\plotone{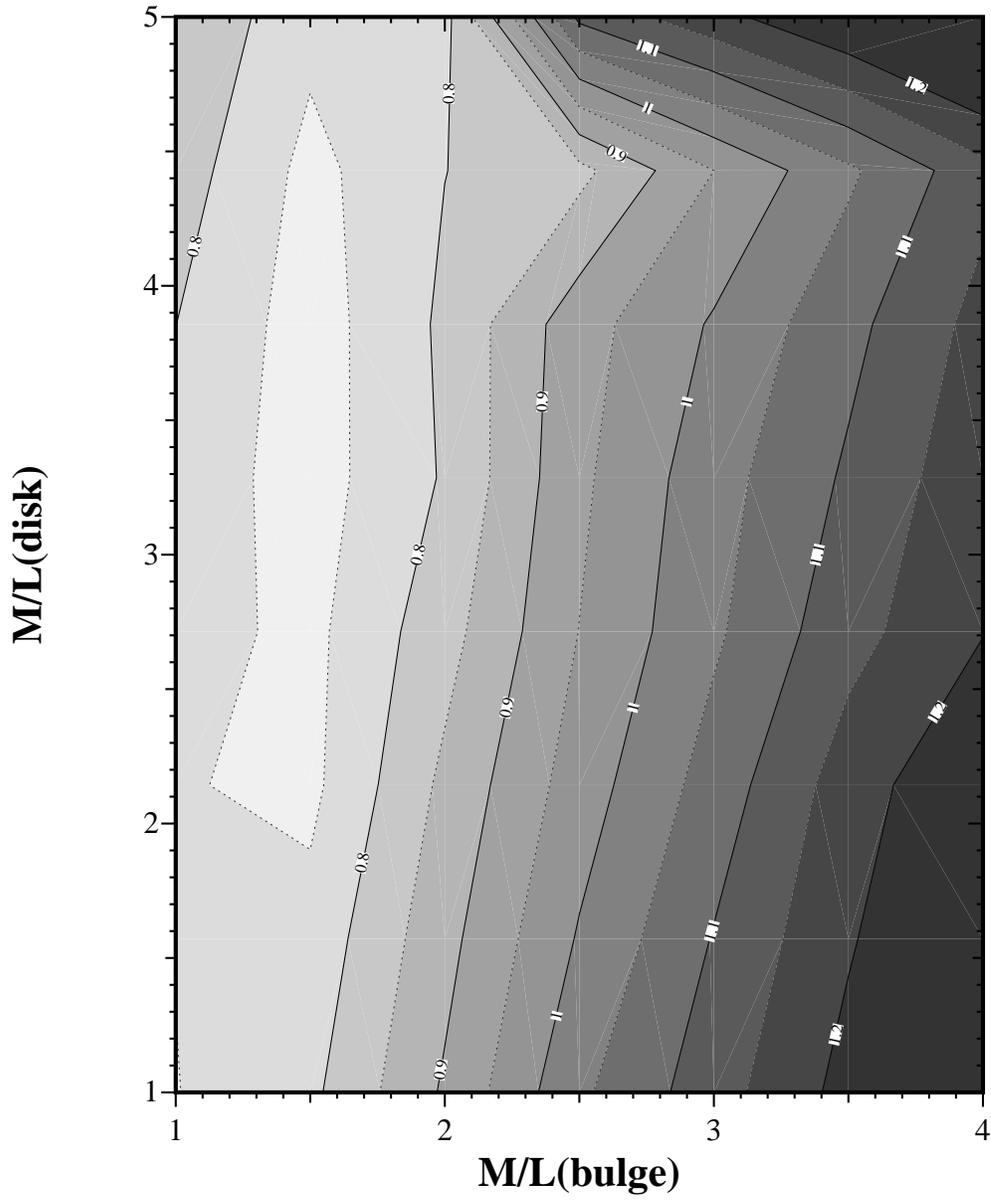} \figcaption{Contour plot of $\chi^2$
in the $\mld-\mlb$ plane. \label{fig:contour}}
\end{center}
\end{figure}

A wide range of models provide acceptable fits to the observations.
Figure \ref{fig:contour} is a contour plot of $\chi^2$ in the
$\mld-\mlb$ plane ($\mld$ and $\mlb$ are the R-band mass-to-light
ratios of the disk and bulge respectively).  The general structure of
a valley running approximately parallel to the $\mld$-axis arises for
two reasons.  First, the bulge luminosity is constrained by the inner
part of the surface brightness profile while its gravitational
potential is constrained by the dispersion data.  Hence, $\mlb$ is
relatively well-determined.  By contrast, the primary constraint on
the disk mass comes from the rotation curve.  But since the disk and
halo contributions to the rotation curve are similar, one can be
played off the other and therefore the $\mld$ is poorly constrained.

The degeneracy with respect to the disk mass-to-light ratio is generic
to modeling of spiral galaxies and has been known for some time (see,
for example, \citet{alb85}).  This degeneracy may be broken by
fixing the disk mass-to-light ratio to a theoretically preferred
value, e.g., one derived from population synthesis models.
Alternatively, one may constrain the disk mass-to-light ratio by
requiring that the galaxy model be stable against the formation of a
strong bar.  As with the Milky Way, M31 may have a weak bar and so
absolute stability is not a requirement (or even desirable).  However,
models with very heavy disks can clearly be ruled out as we
demonstrate in the next section.  A third possibility is to include an
additional data set that probes more directly the disk mass
distribution such as disk velocity dispersion measurements as in the
study by \citet{bot87} of NGC5170.

The mass-to-light ratios in Figure \ref{fig:contour} must be corrected
for foreground and external extinction if we are to make contact with
theoretical predictions.  Foreground extinction toward M31 is estimated
to be $0.41$ mag in B \citep{deV76} corresponding to $0.23$ mag in R
assuming the standard interstellar extinction law \citep{bin98}.
Estimates of (R-band) internal extinction for the disk range from
$0.6$ \citep{mon77} to $0.74$ mag \citep{ken89} while formulae in
\citet{tul98} give $0.64$ mag.  Assuming $0.65$ mag internal
extinction for the disk and no internal extinction for the bulge
\citep{ken89}, $\mld$ and $\mlb$ in Figure \ref{fig:contour} should be
scaled downward by 2.2 and 1.2 respectively to give $\chi^2$ in terms
of intrinsic mass-to-light ratios.

Population synthesis models provide an independent means of
constraining the mass-to-light ratios in disk galaxies.  \citet{bel01}
provide a table of predicted mass-to-light ratios as a function of
various color parameters.  In particular, they find

\begin{equation}\label{eq:belldejong}
\log_{10}\left (M/L_R\right ) = 
-0.820 + 0.851\left (B-R\right )~.
\end{equation}

\noindent For the M31 disk, \citet{wal88} find $B-R\simeq 1.53$ which,
when corrected for extinction, corresponds to $B-R\simeq 1.18$
implying $\mld\simeq 1.5$ (corrected) or $\mld\simeq 3.4$
(uncorrected).  Applying the same formula to the bulge region where
$B-R\simeq 1.6$ yields $\mlb\simeq 3.2$ (corrected) or $3.9$
(uncorrected).

Our results may be compared with those from previous investigations.
\citet{ken89} modeled the disk and bulge of M31 using rotation curve
and bulge dispersion measurements as well as photometric observations
carried out in the r-bandpass of the \citet{thu76} system.  He found
(uncorrected) values for the mass-to-light ratios of $\left
(M/L_r\right )_{\rm disk} =10.5$ and $\left (M/L_r\right )_{\rm
bulge}=5.0$.  Using the transformation $r-R=0.43 + 0.15\left
(B-V\right )$ from \citet{ken85} with a $B-V$ color from \citet{wal88}
Kent's values become $\mld\simeq 6.2$ and $\mlb\simeq 3.0$.  Note that
\citet{ken89} assumes a constant density halo (i.e., halo with a very
large core radius) which explains why his value for $\left ( M/L\right
)_d$ is higher than ours and others \citep{kly02}.

Recent interest in bulge models has been driven in large part by an
attempt to understand their connection with galactic supermassive
black holes.  \citet{mag98} have constructed dynamical models for
bulge-black hole systems in 36 spiral galaxies.  For M31, they find a
mass-to-light ratio in V-band of $4.83\pm 0.1$ or $\mlb\simeq 3.4$
(uncorrected for extinction).

To summarize, our analysis alone does little to constrain the disk
mass-to-light ratio.  As for the bulge, previous modeling as well as
predictions from stellar population studies point to a mass-to-light
ratio a factor of two higher than our preferred value.

\begin{figure}[p]
\begin{center}
\plotone{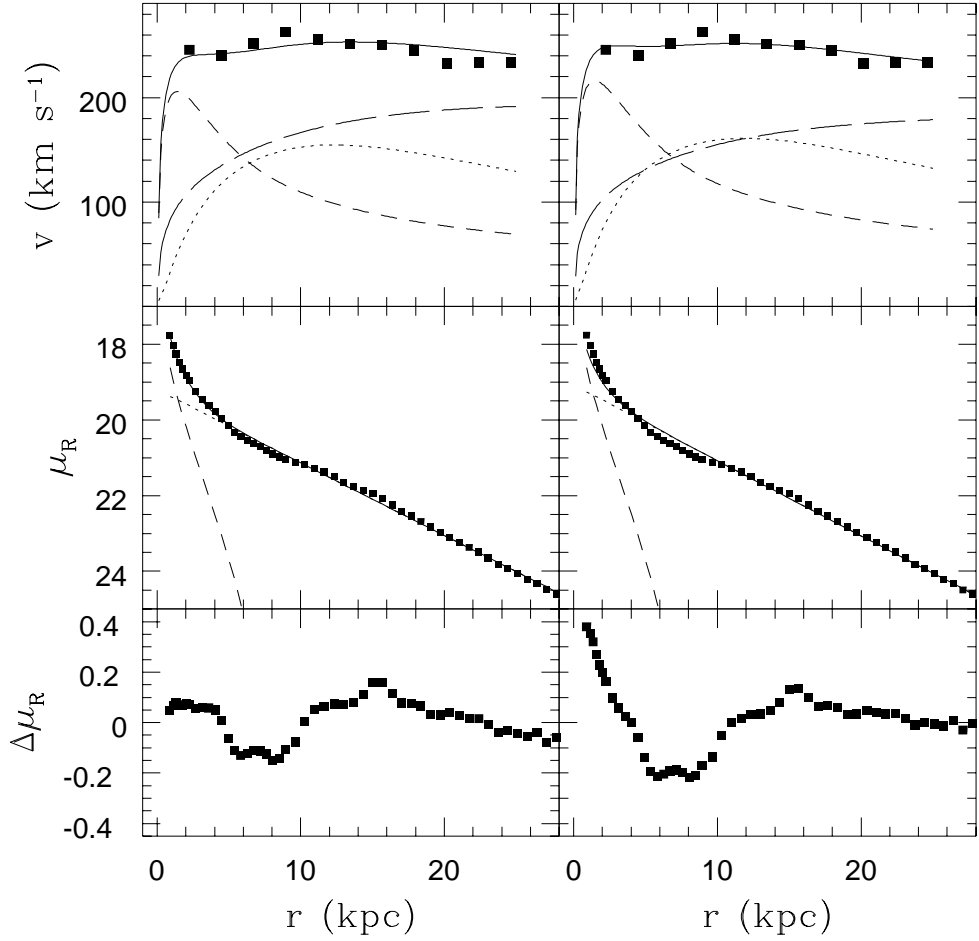} \figcaption{Predictions vs. observations
for models M31a (left) and M31b (right).  Shown are the rotation curve (top),
R-band surface brightness profile (middle) and residuals 
(predicted - observed) for surface
brightness profile (bottom).  Linetypes in the rotation curve
plots are the same as in previous figures.\label{fig:m31a-b-rot-sbp}}
\end{center}
\end{figure}

\begin{figure}[p]
\begin{center}
\plotone{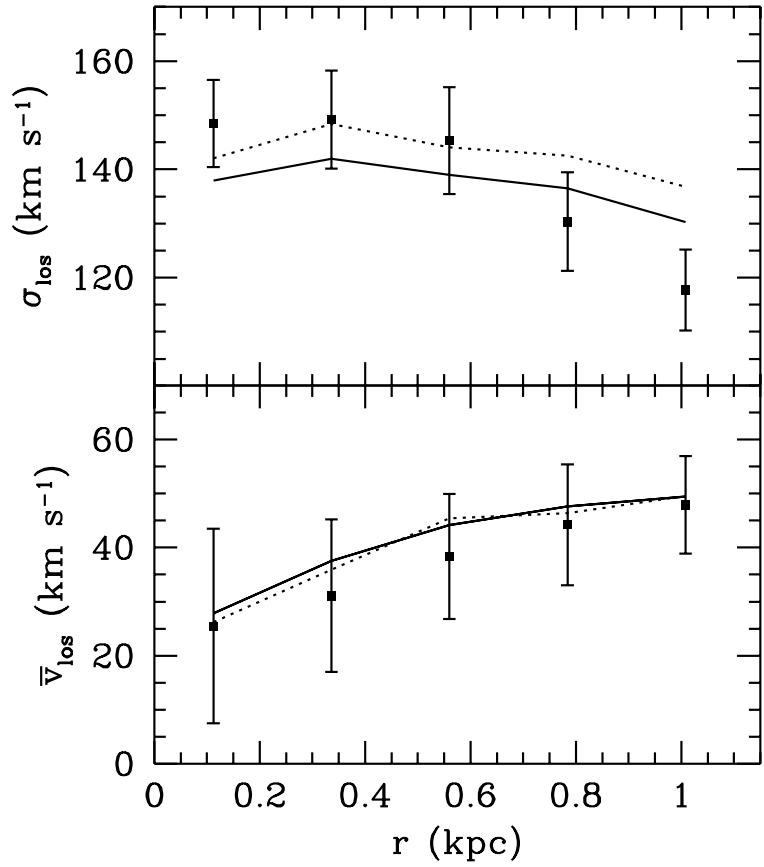} \figcaption{Bulge dispersion
and bulk rotation profiles for models M31a (solid curve), M31b (dotted curve) 
and observations (data points).\label{fig:m31-bulgedispersion}}
\end{center}
\end{figure}

To better understand these results, we consider four models in more
detail.  Model M31a has disk and bulge mass-to-light ratios set equal
to one another and to the value predicted by population synthesis
models.  This model lies close to the $\chi^2$-valley in Figure
\ref{fig:contour}.  The same disk mass-to-light ratio is used in M31b
but here $\mlb$ is increased to $3.4$ as estimated by \citet{mag98}.
The rotation curve and surface brightness profiles for these two
models are shown in Figure \ref{fig:m31a-b-rot-sbp} while the bulge
line-of-sight dispersion and bulk velocity profiles are shown in
Figure \ref{fig:m31-bulgedispersion}.  The preference for lower values
of $\mlb$ is evident in the surface brightness and bulge line-of-sight
dispersion profiles.  With the higher value of $\mlb$, the predicted
dispersion profile is too flat, i.e., does not fall fast enough with
radius.  The simplex algorithm strikes a balance between a bulge that
is too dim and one that is too massive but ultimately, cannot
fit the surface brightness and dispersion profiles as well as the
low-$\mlb$ model.  Note that \citet{ken89} has the same difficulty
with the bulge dispersion profile while the analysis of \citet{mag98}
is restricted to the innermost $0.15\,{\rm arcmin}\,(30\,{\rm pc})$ of
the bulge.

\begin{figure}[p]
\begin{center}
\plotone{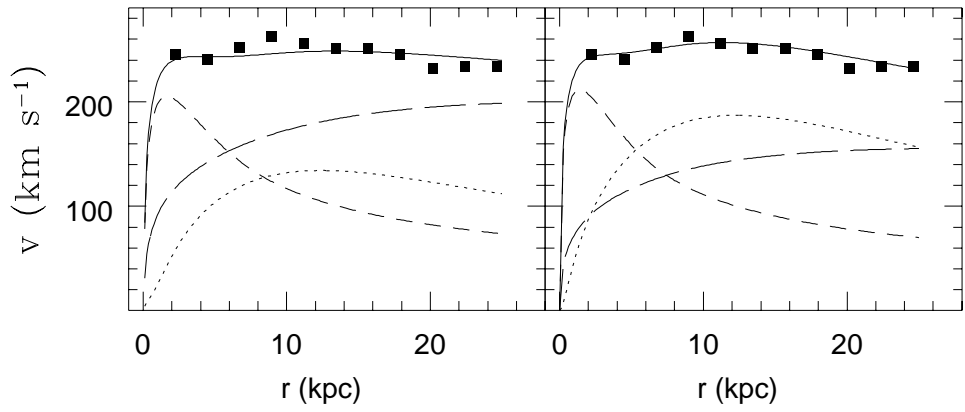} \figcaption{Rotation curves for models M31c (left)
and M31d (right).\label{figure-d}}
\end{center}
\end{figure}

M31c-M31a-M31d form a sequence of models from light to heavy disk
mass.  The rotation curves for models M31c and M31d are shown in
Figure \ref{figure-d}.

\subsection{Equilibrium and Stability}

We simulated N-body realizations of the four M31 models to test for
both the quality of the derived equilibria and stability against bar
formation.  In all four models, the bulge dominates the innermost part
of the rotation curve.  In M31c, the disk contribution is subdominant
throughout the system whereas in model M31d, the disk dominates the
rotation curve between $5$ and $25\,{\rm kpc}$.  We anticipate that
M31d will form a bar while M31c will be stable.  For models M31a and
M31b the disk contribution to the rotation curve is comparable to that
of the other components at about one scale radius.

For each model, we generate an N-body realization containing a total
of 3.5M particles (1M disk particles, 500K bulge particles and 2M halo
particles) and evolve the system using a parallel N-body treecode
\citep{dub98b}.  The extent to which the initial conditions represent
a system in equilibrium and its susceptibility to the bar instability
are explored by monitoring the surface density of the disk+bulge
system, the disk velocity ellipsoid, the disk scale height, and the
density profiles of the bulge and halo.  We begin by discussing the
results for model M31a which does not form a bar and represents a good
test case.  The central disk radial velocity dispersion has been set
so that the Toomre $Q$ parameter is $Q\approx 1.2$ at $R=2 R_d$.  This
initial disk is rather cool and is unstable to spiral instabilities
arising from the swing amplification of particle shot noise, an effect
that leads to emerging spiral structure and disk heating.
The disk is also heated through its interactions with halo particles
but this effect is relatively minor with $2M$ halo particles and a
halo particle mass is only $m_h=2.9\times 10^5$ M$_\odot$.  (A halo
made of $10^6$ M$_\odot$ black holes will roughly double the disk
scale height over a Hubble time (e.g. \citet{lac85}) and so our halo
should have a much smaller effect.)

\begin{figure}[p]
\begin{center}
\plotone{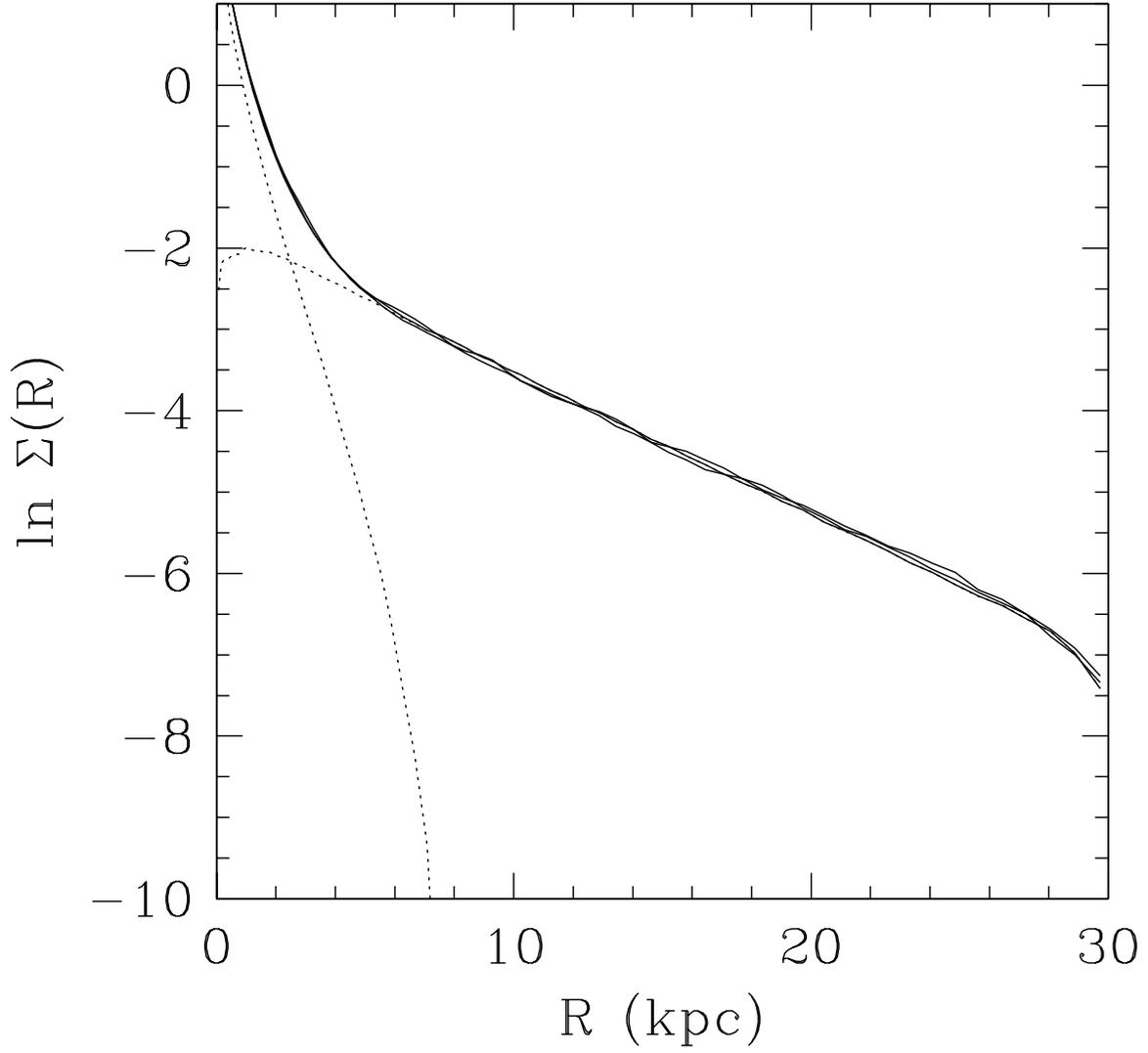}
\figcaption{Evolution of the surface density profile 
over 4.8 Gyr.  The three solid curves give
the total surface density at $t=0,\,2.4,\,$ and $4.8\,{\rm Gyr}$.
The dotted curves show the separate contributions from the bulge and disk.
\label{fig:sden}}
\end{center}
\end{figure}

\begin{figure}[p]
\begin{center}
\plotone{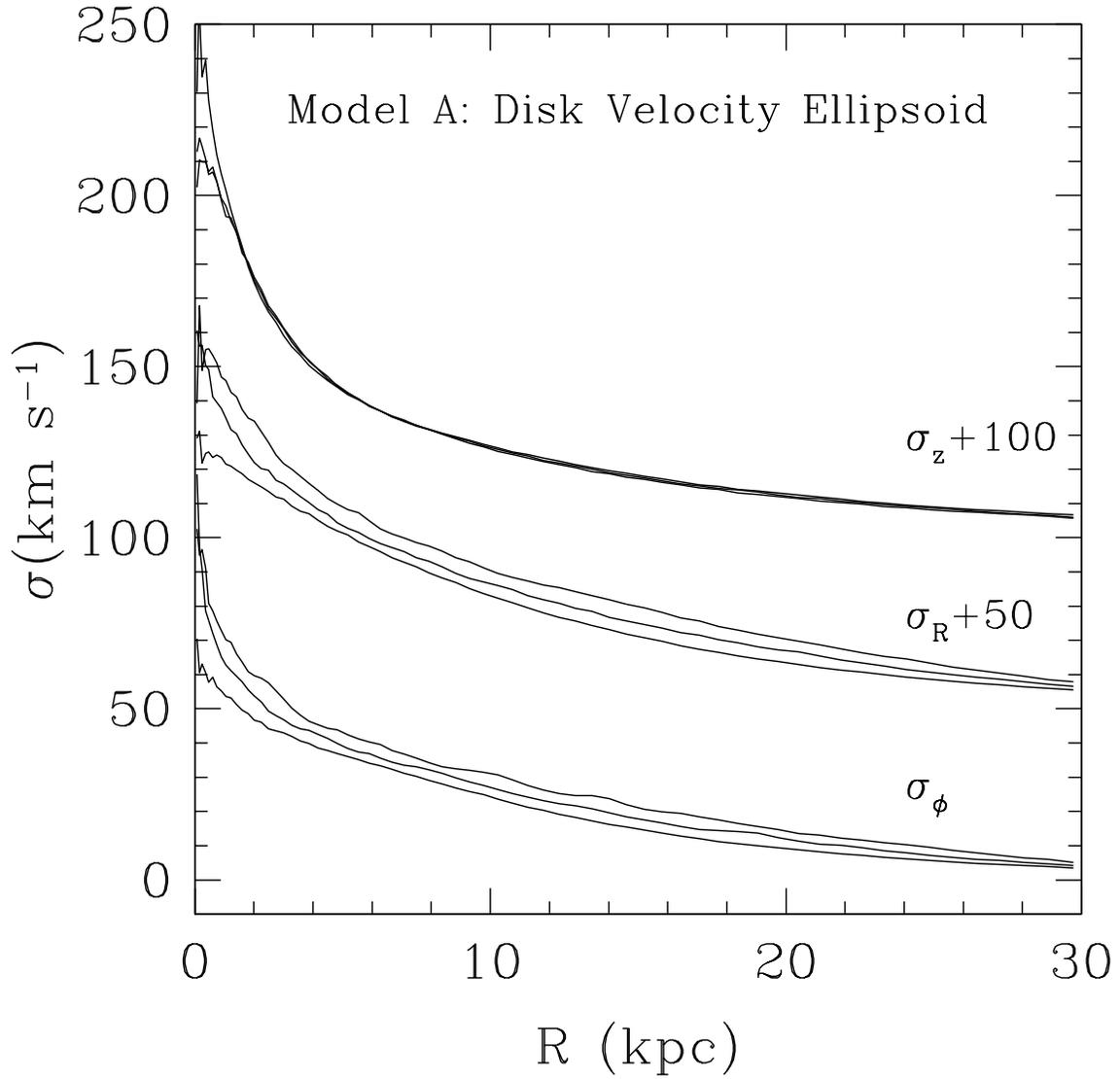}
\figcaption{Evolution of the disk velocity ellipsoid over 10 Gyrs.  The profiles of the velocity dispersions are shown at $t=0.0, 2.4$ and 4.8 Gyr showing a slow increase over the disk.\label{fig:disk-vsig}}
\end{center}
\end{figure}

\begin{figure}[p]
\begin{center}
\plotone{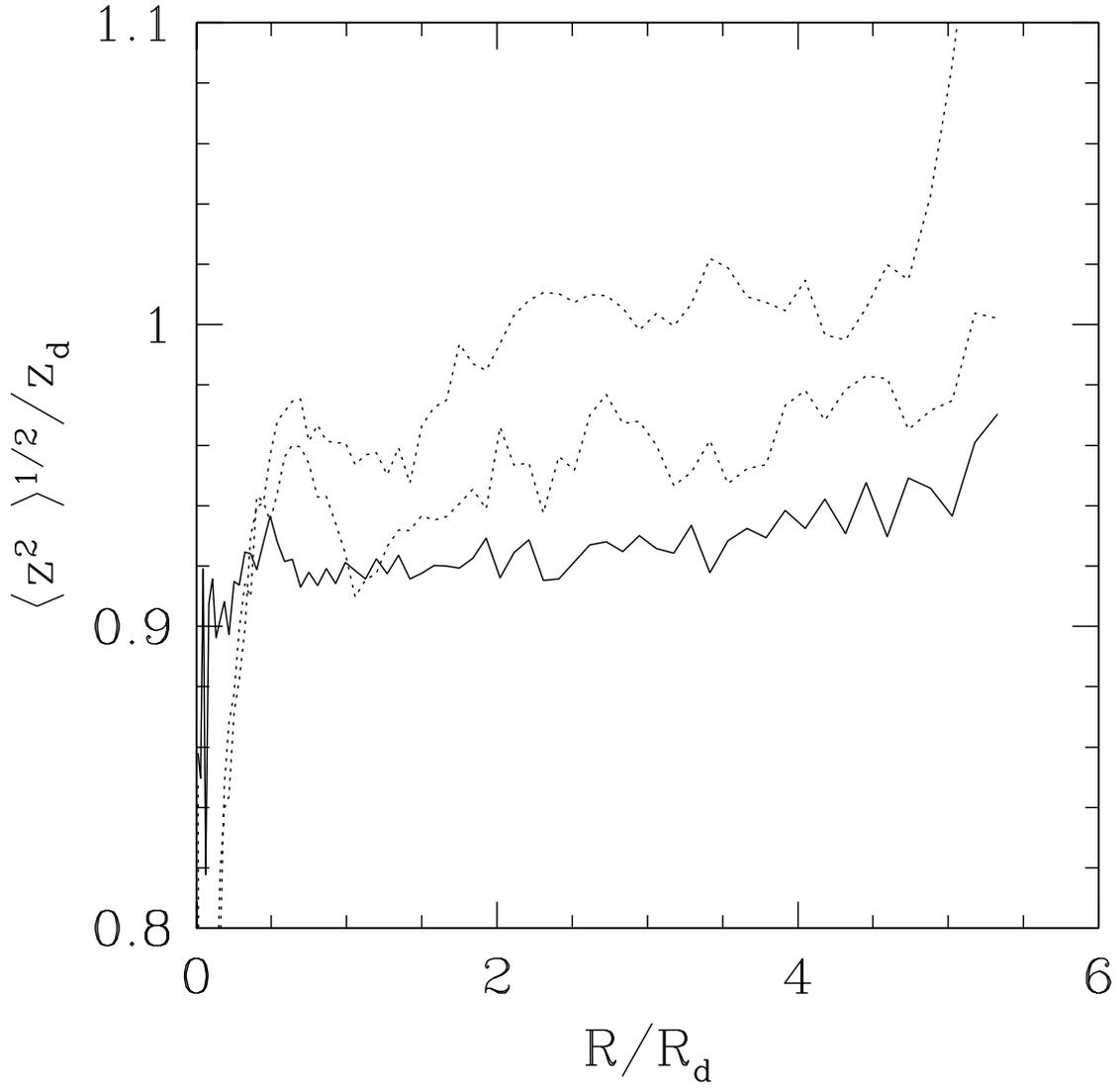}
\figcaption{Disk scale height evolution at $t=0$ (solid curve)
and $t=2.4$ Gyr and $t=4.8$ Gyr (dotted curves)\label{fig:zh}}
\end{center}
\end{figure}

We ran Model M31a for $t=4.8$ Gyr using 10000 timesteps.  The particle
softening radius was set to $s=50$ pc so the number of timesteps was
more than adequate to follow orbits down to the softening radius in
the galaxy core.  The first impression is that the initial conditions
are almost in perfect equilibrium with essentially no transient
readjustment of the disk at start-up, i.e., no evidence of the
imperfections in the initial conditions that are present in other
methods (e.g. \citet{her93}).  Figure \ref{fig:sden} shows the
disk+bulge surface density profile at $t=0,2.4$ and 4.8 Gyr along with
the disk and bulge initial profiles.  There is no change in the
profile over the 4.8 Gyr integration and there is no sign of a bar.
Figure \ref{fig:disk-vsig} shows the evolution of the radially
averaged disk velocity ellipsoid throughout the disk and reveals the
effect of natural heating through the growth of spiral structure.  The
disk velocity ellipsoid at $R=15\,{\rm kpc}$ (about 3 disk exponential
scale lengths) changes from $(\sigma_R,\sigma_\phi,\sigma_z) =
(22,15,18)$ km s$^{-1}$ to $(31,22,19)$ km s$^{-1}$ after 4.8 Gyr.
Another indicator of disk evolution and heating is the vertical
scale-height.  Figure \ref{fig:zh} shows the variance in disk particle
height above the midplane as an estimator of disk scale-height at 3
times.  There is some evolution with the scale-height growing by about
10\% showing the good quality and relatively low shot noise effect in
the models.  The apparent flaring beyond about $R=4R_d$ may result
partially from poor vertical force resolution due to discrete sampling
of the disk.  Since we choose a constant disk particle mass, the
particle number density drops off quickly at large radii.

\begin{figure}[p]
\begin{center}
\plotone{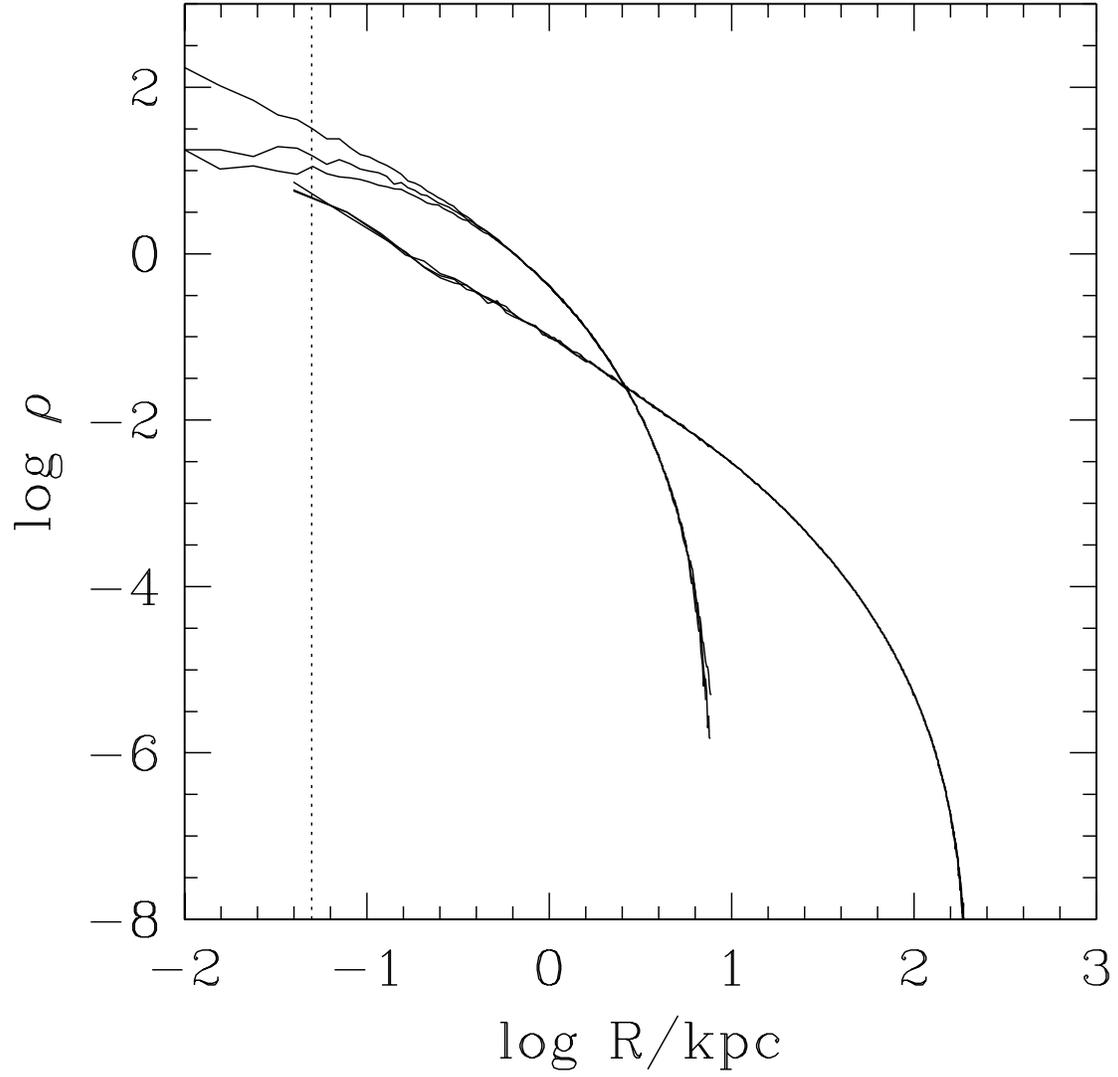} \figcaption{Bulge and halo radial density
profiles over 4.8 Gyr.  The vertical dotted line corresponds to the
softening length.  A constant density core develops in the bulge on the
scale of the softening length of $s=50$ pc.\label{fig:hbden}}
\end{center}
\end{figure}

\begin{figure}[p]
\begin{center}
\plotone{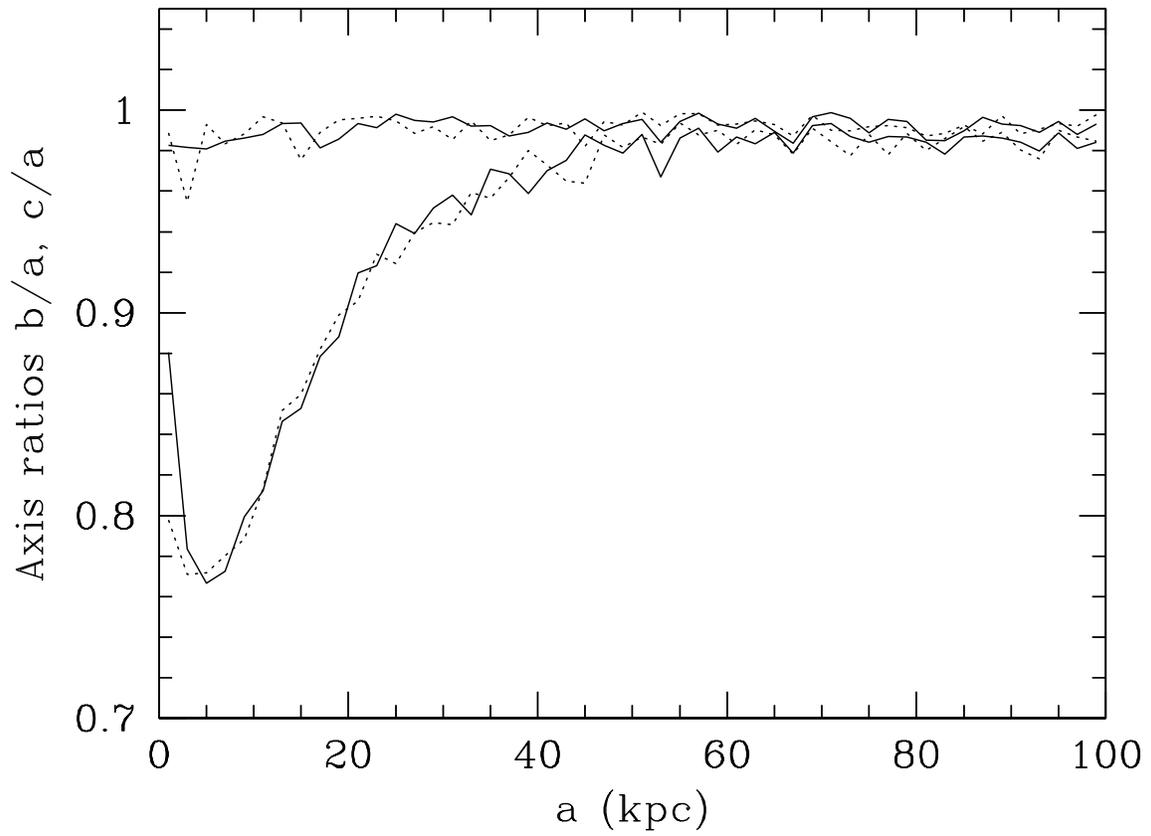} \figcaption{Halo axis ratio profile at
the start (solid) and end (dashed) of the simulation over 4.8 Gyr.
With the exception of the central point the axis ratio of the halo
remains essentially constant throughout the simulation.
\label{fig:shapeprofile}}
\end{center}
\end{figure}

Figure \ref{fig:hbden} shows the density profile of the bulge and halo
over the course of the simulation.  The denser bulge develops a core
radius of $r_c\approx 300\,{\rm kpc}$ because of the use of force
softening while the halo maintains its $r^{-1}$ cusp.  We see below in
models with black holes that the cusp profile can be maintained if
required using smaller softening lengths and smaller timesteps.  We
also examined the evolution of the halo shape profile as a further
test.  Near the vicinity of the disk, the halo profile shape is
flattened by the presence of the disk potential and so that flattening
should remain unchanged if the system is in proper equilibrium.  We
compute the shape of the best fit ellipsoidal density contours as a
function of radius at the start and end of the simulation using the
normalized inertial tensor using the method described in Dubinski and
Carlberg (1991) (Figure \ref{fig:shapeprofile}).  This shows that the
halo is flattened into an oblate spheroid with $q\simeq 0.8$ near the
disk and then becomes spherical at larger radii.  With the exception
of the central point where softening effects modify profiles, the
shape profile remains unchanged throughout the simulation implying a
good choice of equilibrium.  In summary, the method produces an
excellent equilibrium configuration of a spiral galaxy.  The gradual
heating of the disk can be attributed to the formation of transient
spiral structure and to a lesser extent heating by halo particles.

Simulations of the remaining models also show clean initial
equilibria.  Only model M31d develops a bar.  The bar forms at
$3\,{\rm Gyr}$ and persists until the end of the simulation at $t=4.8$ Gyr.
At the end of the simulation the bar has a length of $r_{b} =
7.7$ kpc with a pattern speed of 23 km s$^{-1}$ kpc$^{-1}$ and a
co-rotation radius of $D_L=9.8$ kpc.  The ratio $D_L:r_b=1.3$ makes
this a ''fast" bar in the standard nomenclature (e.g. \citep{one03})
at least at this stage of the simulation.  While M31 appears to have a
triaxial bar-like bulge, at present, it does not appear to be a
well-developed barred spiral.  The absence of a bar in M31 suggests
that the disk mass in Model D is too great and that one of the other
models is more acceptable for M31.

We also simulated the two Milky Way models and found little
evolution in the velocity ellipsoid, vertical scale height, surface
density, and space density again illustrating the excellent quality of
the models as a means of setting up initial conditions for N-body
simulations.  Not surprisingly, model MWa (where the disk provides the
dominant contribution to the rotation curve at intermediate radii)
develops a bar while MWb appears to be stable.

\subsection{Incorporating a black hole}

In this section, we examine an axisymmetric model of M31
augmented by a central black hole.  M31 is the nearest galaxy with a
demonstrable central black hole but unfortunately exhibits a complex
central structure in the form of a double nucleus \citep{lau93,
bac94}.  Estimates of the black hole mass based on dynamical models
and the observed bulge surface brightness and velocity dispersion
profiles fall in the range $3 - 8.5\times 10^{7}\,M_\odot$
\citep{tre95, kor99, bac01}.

\begin{figure}[p]
\begin{center}
\plotone{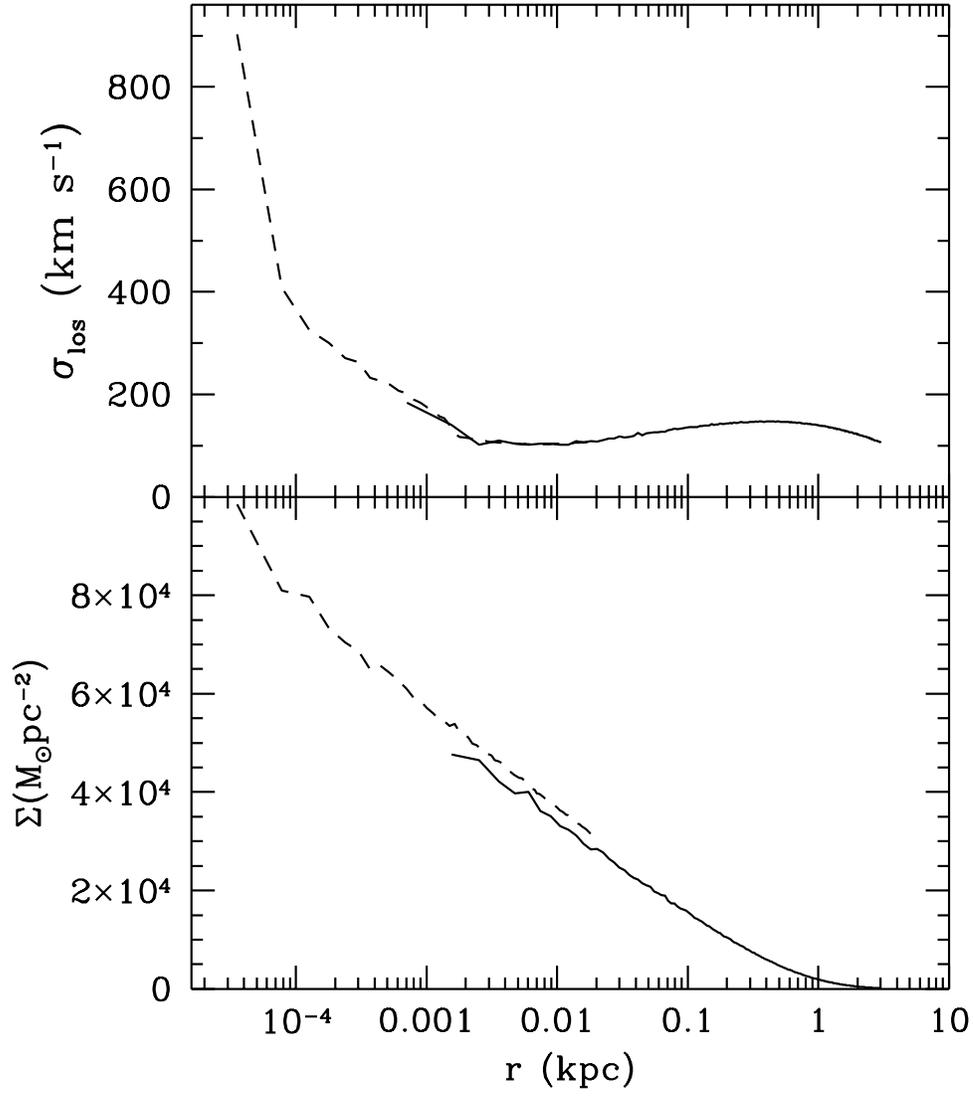} \figcaption{Velocity dispersion (top) and surface
brightness (bottom) profiles in the bulge.
\label{fig:m31bh1}}
\end{center}
\end{figure}

\begin{figure}[p]
\plotone{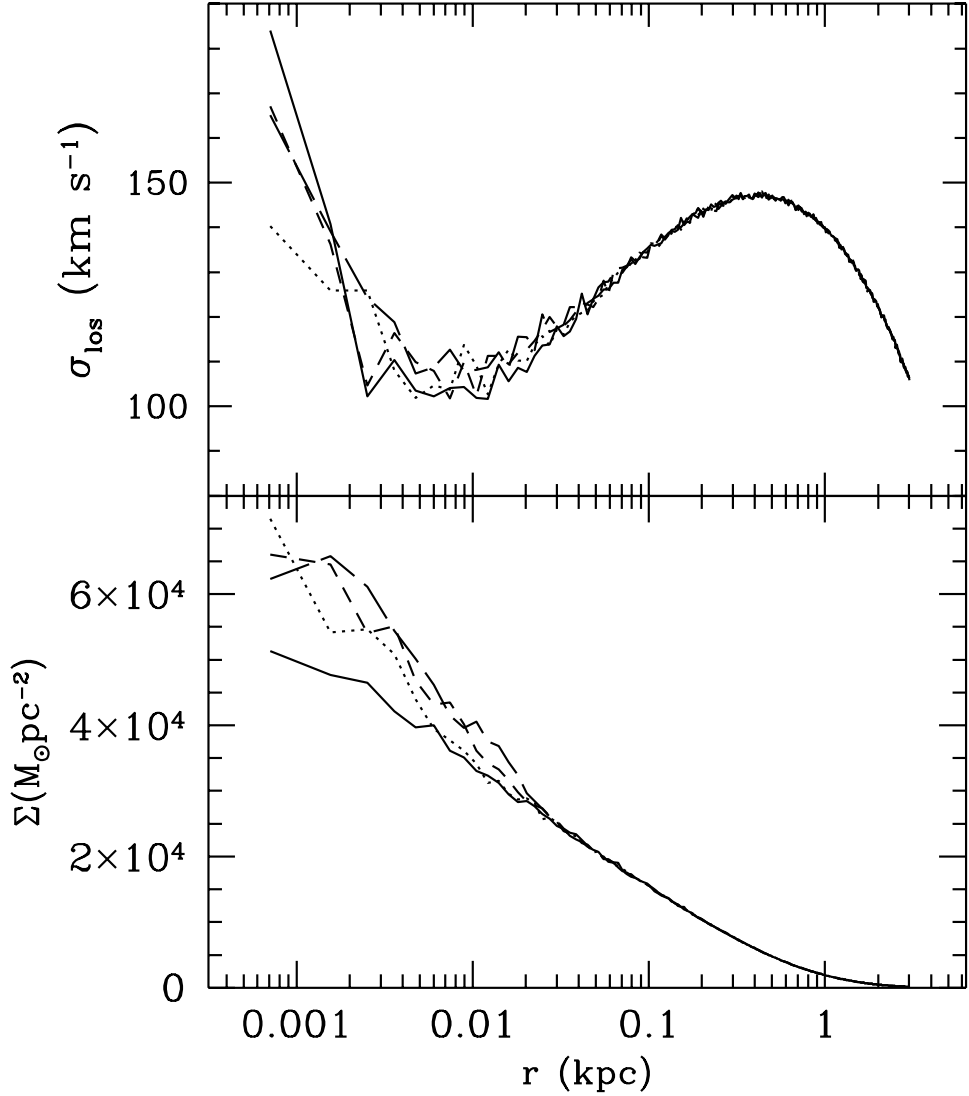} \figcaption{Evolution of the velocity dispersion and
surface brightness profiles around the central black hole.  Solid
curve -- $t=0$; dotted curve -- $t\simeq 10^5\,{\rm yrs}$ short-dashed
curve -- $t\simeq 2\times 10^5\,{\rm yrs}$ long-dashed curve --
$t\simeq 3\times 10^5\,{\rm yrs}$.  The orbital time for the black
hole at $1\,{\rm pc}$ is $1.7\times 10^4\,{\rm yrs}$.
\label{fig:m31bh2}}
\end{figure}

As an illustration of our models, we place a black hole of mass $3\times
10^7\,M_\odot$ at the center of model M31a.  Figures \ref{fig:m31bh1}
and \ref{fig:m31bh2} show the line-of-sight velocity dispersion and
surface density profiles.  To generate the inner portion of the curves
(dashed lines) in Figure \ref{fig:m31bh1} we sample 10M particles from
the bulge DF within a thin tube 20 pc in radius centered on the galaxy
and aligned with the symmetry axis.  The dispersion profile exhibits
the characteristic intermediate minimum and maximum seen in the data
\citep{tre95}.

We also simulate a galaxy model including a blackhole to test the
quality of the equilibrium focusing on the nuclear region close to the
blackhole.  The blackhole moves freely as single particle within the
simulation.  For this test, we put most of the particles in the bulge,
$N=10M$ so the ratio of the blackhole mass to the bulge particle mass
is $\sim 10^4$.  The enclosed mass of bulge stars equals the blackhole
mass at a radius $r=34\,{\rm pc}$.  there are only The halo and disk
are represented by 2M and 1M particles respectively but their
particles are unlikely to interact with the blackhole in this test.
We set the particle and blackhole softening length to $s=1\,{\rm pc}$
and use a single timestep of $\Delta t=100\,{\rm yrs}$.  Even with a
total of $N=10M$ particles in the bulge, there are only 13 bulge
particles initially within the softening radius.  Figure
\ref{fig:m31bh2} also shows the evolution of the model focusing on the
central region surrounding the black hole.  The simulation is run for
3000 timesteps corresponding to approximately 18 orbital times at a
radius of $1\,{\rm pc}$.  We see that there is some evolution of the
system --- towards the center, the velocity dispersion fluctuates
slightly within $r\sim 10$ pc and the surface density increases
slightly.  At $r=1$ pc where there are only a dozen or so particles,
we expect larger initial variations that may explain the settling to a
slightly higher density.  Nevertheless, the profiles remain close to
the initial state suggesting that the equilibrium is reasonable.
These effects may be also result indirectly from using force softening
or a reflection of the approximations that have gone into the DF.
Clearly, high numerical resolution is needed to treat the dynamics of
stars around a central blackhole and the short orbital times make
these technically difficult and costly.

Models such as this one may be used as initial conditions for
numerical experiments of black hole dynamics during galaxy merger
events such as the simulations by \citet{mil03} and will be explored
in future investigations.

\section{Summary and Discussion}

Our goals in this paper have been threefold.  First we have presented
a new set of model DFs for multi-component disk galaxies.  Second, we
have identified particular models that fit observational data for the
Milky Way and M31.  Finally, we have explored the stability of the
models using numerical simulations.

Our DFs respresent self-consistent axisymmetric equilibrium solutions
to the Poisson and CB equations.  The disk and bulge DFs are motivated
by observations while the halo DF is motivated by results from
cosmological N-body simulations.  The models permit the inclusion
of a central supermassive black hole.

Historically, two approaches have been used to construct DFs for
equilibrium systems \citep{bin87}.  The first approach is to propose a
DF built from chosen functions of the integrals of motion and then
calculate the density by solving the Poisson equation. The second is
to calculate the DF from the desired density profile via an integral
equation known as the Abel transform.  Our DFs are the product of a
hybrid scheme.  The DF for the isolated NFW halo was derived from the
Abel transform, modified by the energy mapping as described in the
text, and then used for the halo in the composite system.

Our models are defined by 15 free parameters which may be tuned to fit
a wide range of observational data.  Typically most of the parameters
are poorly constrained.  A galaxy's surface brightness profile is, in
general, adequate to fix the disk scale length and, in combination
with velocity dispersion and rotation curve measurements, provides a
constraint on the disk mass and halo and bulge structural parameters.
However degeneracies remain primarily due to uncertainties in the disk
and bulge mass-to-light ratios.  Theoretical considerations such as
results from population synthesis studies or numerical experiments of
the bar instability can help narrow down the field of acceptable
models.  Additional data for external galaxies such as velocity
dispersion measurements in the disk may help break the $M/L$
degeneracy.  Observations of edge-on spiral galaxies can be used to
constrain the vertical scale height and disk trucation parameters
\citep{kre02}.

We have constructed sequences of models for the Milky Way and M31
which provide excellent fits to available data.  The models serve
to illustrate the general proceduce for searching a large parameter
space to find acceptable models for particular galaxies.

A primary purpose of our models is to provide initial conditions for
N-body experiments and well-developed techniques allow one to sample
the DFs with arbitrary numbers of particles.  Through a series of
numerical experiments, we have explored the quality of the models as
initial conditions and their stability to the formation of bars.
Models in which the disk contribution to the rotation curve is always
subdominant to that of the bulge and halo tend to be stable against
bar formation and therefore provide the best laboratory to study the
quality of the model DFs.  Our analysis of one such model indicates
that the DFs do indeed provide excellent initial conditions.  The
radial profiles of the space density, surface density, velocity
dispersion tensor, and vertical scale height all remain relatively
constant over $4.8\,{\rm Gyr}$.  Areas where we do see some evolution
may be understood from simple considerations.  For example, the bulge
evolves from Hernquist cusp to core inside the softening length
because the velocity distribution is calculated assuming an unsoftened
force-law.  Spiral density waves, generated by swing-amplification of
shot noise cause the disk to thicken, but only by about 10\%.

Models in which the disk contribution to the rotation curve is
dominant over some range in radius, develop strong bars.  We confirm
this well-known result for both Milky Way and M31 models.

Our models allow for the addition of a central supermassive black
hole.  Our prescription for doing this maintains the density
distribution of the black hole-less models while modifying the
velocities to establish a new dynamical equilibria near the black
hole.  The velocity dispersion profiles of our black hole models have
the right characteristics to match the data, namely high dispersion
near the black hole falling to a minimum at tens of parsecs, rising at
a radius of a hundred or so parsecs, and then falling again as one
moves further out in radius.  Thus, we may be able to model galaxies
from the sphere of influence of the black hole out to the virial
radius, an impressive dynamic range of five or more orders of
magnitude.

There are numerous open avenues for future work.  The model fitting
algorithm can accomodate any number and type of observations and one
may add to the pseudo-observations layers of realism such as
statistical fluctuations and the effects of seeing.  Our M31 study
used a small subset of the available photometric and kinematic data.
Two-dimensional surface brightness and dispersion maps reveal the
presence of the bar and may be compared with corresponding maps from
the ``evolved'' models providing a further constraint on the model.

A classic problem in galactic dynamics is the disruption of satellite
systems by the tidal field of the parent galaxy.  The discovery of an
arc of stars associated with the Sagittarius dwarf galaxy (see
\citet{maj03} and references therein) has intensified interest in
performing simulations of the tidal disruption of this system
\citep{iba01,hel04} (see also \citet{gee05} for a study of the
Andromeda stream in a disk-bulge-halo model of M31).  Our models offer
the possibility of performing such simulations realistic and
fully-self-consistent models for both parent and satellite systems.
Since a primary goal of previous investigations has been to constrain
the shape of the Galactic halo, our models will have to be extended to
include triaxial systems if they are to be useful in this endeavor.
The extension to triaxial systems might be accomplished by
adiabatically molding the models as in \citet{hol01}.

Our models might be extended in other important ways such as the
inclusion of a globular cluster system or thick disk as discussed in
the text.  Perhaps the most significant and challenging improvement
would be to add a gas component and star formation since this would
enable the study of spiral structure and bar formation with a far
greater level of realism.

\acknowledgements{It is a pleasure to thank S. Courteau, R. Henriksen,
K. Perrett, and S. Kazantzidis for useful conversations.  This work
was supported, in part, by the Natural Science and Engineering
Research Council of Canada and the Canadian Foundation for
Innovation.}


\newpage

\begin{deluxetable}{cc}
\tablewidth{0pt}
\tabletypesize{\footnotesize}
\tablecaption{summary of model parameters}
\tablehead{ \colhead{parameter} & {description} \\}
\startdata
$\epsilon_h$ & halo tidal radius parameter \\
$\sigma_h$ & halo characteristic velocity \\
$a_h$	& halo scale length \\
$\alpha_h$ & halo bulk rotation \\
$M_d$ & disk mass  \\
$R_d$ & disk scale length \\
$R_{\rm out}$ & disk truncation radius \\
$\delta R_{\rm out}$ & sharpness of truncation \\
$h_d$ & disk scale height \\
$\sigma_{R0}$ & radial velocity dispersion at galaxy center \\
$R_\sigma$ & scale length for radial dispersion \\
$\epsilon_b$ & bulge tidal radius parameter \\
$\sigma_b$ & characteristic bulge velocity \\
$a_b$ & bulge scale length \\
$\alpha_b$ & bulge rotation
\enddata
\end{deluxetable}

\begin{deluxetable}{ccccccccccccc}
\tablewidth{0pt}
\tabletypesize{\footnotesize}
\tablecaption{parameters for models discussed in the text\label{tab:models}}
\tablehead{ \colhead{Model} & \colhead{$\epsilon_h$} &
 \colhead{$\sigma_h$} & \colhead{$a_h$} &
 \colhead{$M_d$} & \colhead{$R_d$} & \colhead{$h_d$} & 
 \colhead{$\epsilon_b$} & \colhead{$\sigma_b$} &\colhead{$a_b$} & 
 \colhead{$\sigma_{R0}/\omega_b$} & \colhead{$\mld$} & \colhead{$\mlb$} \\
}
\startdata
Sample & 0.079 & 1 & 1 & 0.1 & 0.3 & 0.02 & .1 & 1.15 & 0.15 & -- & -- & --\\
MWa & 0.17  & 2.496  & 12.96 & 19.66 &  2.806 & 0.409
& 0.787  & 4.444  & 0.788 & 1.211 & -- & -- \\
MWb & 0.11 & 3.447 & 8.818 & 14.47 & 2.817 & 0.439 & 0.791 & 4.357 & 
0.884 & 1.244 & -- & --\\
M31a & 0.25 & 3.371 & 12.94 & 33.40 & 5.577 & 0.3 & 0.929 & 4.607 & 1.826 
& 0.763 & 3.4 & 1.9 \\
M31b & 0.25 & 3.224 & 14.03 & 35.08 & 5.401 & 0.3 & 0.925 & 4.811 & 1.857 
& 0.751 & 3.4 & 3.4 \\
M31d & 0.36 & 3.243 & 17.46 & 50.10 & 5.566 & 0.3 & 0.924 & 4.685 & 1.802
& 0.732 & 5.0 & 2.5 \\
\enddata
\tablecomments{We assume $G=1$.  Units for the Milky Way and M31
models are ${\rm kpc},~100\,{\rm km\,s^{-1}}$ and $2.33\times 10^9\,M_\odot$.}
\end{deluxetable}

\begin{deluxetable}{cccccccccc}
\tablewidth{0pt}
\tabletypesize{\footnotesize}
\tablecaption{Comparison of observations and model predictions
\label{tab:results}}
\tablehead{
 \colhead{} & \colhead{$\chi^2$} & 
 \colhead{$A$} & \colhead{$B$} &
 \colhead{$\overline{v_R^2}$} & \colhead{$\overline{\Delta v_\phi^2}$} &
 \colhead{$\overline{v_z^2}$} & 
 \colhead{$K_z/2\pi G$} & \colhead{$\sigma_{\rm los}(210\,{\rm pc})$}
 & \colhead{$M(100\,{\rm kpc})$}
}
\startdata
observed  && 14.5  & -12.5  & 36   & 25   & 20    & 71   & 136 & 70 \\
MWa & 0.75  & 13.6  & -12.5  & 31.7 & 26.4 & 19.9  & 71.7  & 134  & 67 \\
MWb  & 0.91 & 13.5  & -12.9  & 32.3 & 27.5 & 20.0  & 68.6 & 134 & 72 \\
\enddata
\tablecomments{Units are ${\rm km\,s}^{-1}$ for the velocity dispersions,
$10^{10}\,M_\odot$ for $M_b$ and $M(100\,{\rm kpc})$ and $M_\odot\,{\rm pc}^{-2}$ for $K_z/2\pi G$.}
\end{deluxetable}

\begin{deluxetable}{cccccccc}
\tablewidth{0pt}
\tabletypesize{\footnotesize}
\tablecaption{derived quantities for Milky Way models
\label{tab:derived}}
\tablehead{
 \colhead{Model} & 
 \colhead{$M_b$} & \colhead{$Q$} &
 \colhead{$K_{z,{\rm vis}}/2\pi G$} & \colhead{$M_t$} &
 \colhead{$r_t$} & 
 \colhead{$c_t$} &
 \colhead{$\rho(R_0)$}

}
\startdata
MWa  & 1.1 & 1.3  & 53 & 74 & 239  & 19  & 0.0079\\
MWb  & 1.2 & 2.2  & 39 & 77 & 236  & 27  & 0.015\\
\enddata

\tablecomments{Units are ${\rm kpc}$ for $r_t$ and $M_\odot\,{\rm
pc}^{-3}$ for $\rho(R_0)$.  Otherwise the units are the same as in
Table 2.}
\end{deluxetable}

\end{document}